\documentclass{amsart}
\usepackage{amsfonts}

\theoremstyle{plain}

\numberwithin{equation}{section}

\input{tcilatex}

\begin{document}
\title[Quantum Nondemolition Principle]{Nondemolition Principle of Quantum
Measurement Theory}
\author{V.P. Belavkin}
\address{Philipps Universit\"{a}t, Fachbereich Physik D--3550, Marburg,
Germany and University of Nottingham, NG7 2RD, UK}
\email{ vpb@maths.nott.ac.uk}
\date{Received August 31, 1992 }
\thanks{ Published in:\emph{\ }\textit{Foundations of Physics}, \textbf{24}
(1994) No 5, 685--714}

\begin{abstract}
\noindent We give an explicit axiomatic formulation of the quantum
measurement theory which is free of the projection postulate. It is based on
the generalized nondemolition principle applicable also to the unsharp,
continuous--spectrum and continuous-in-time observations. The
\textquotedblleft collapsed state--vector\textquotedblright\ after the
\textquotedblleft objectification\textquotedblright\ is simply treated as a
random vector of the \textit{a posteriori}\emph{\/} state given by the
quantum filtering, i.e., the conditioning of the \textit{a priori\/} induced
state on the corresponding reduced algebra. The nonlinear phenomenological
equation of \textquotedblleft continuous spontaneous
localization\textquotedblright\ has been derived from the Schr\"{o}dinger
equation as a case of the quantum filtering equation for the diffusive
nondemolition measurement. The quantum theory of measurement and filtering
suggests also another type of the stochastic equation for the dynamical
theory of continuous reduction, corresponding to the counting nondemolition
measurement, which is more relevant for the quantum experiments.
\end{abstract}

\maketitle

\section{The status of quantum measurement theory}

Quantum measurement theory, based on the ordinary von Neumann or a
generalized reduction postulate, was never an essential part of quantum
physics but rather of metaphysics. First, this was because the orthodox
quantum theory had always dealt with a closed quantum system while the
object of measurement is an open system due to the interaction with the
measurement apparatus. Second, the superposition principle of quantum
mechanics, having dealt with simple algebras of observables, is in
contradiction with the von Neumann projection postulate while it may be not
so in the algebraic quantum theory with the corresponding superselection
rules. Third, due to the dynamical tradition in quantum theory going on from
the deterministic mechanics, the process of the measurement was always
considered by theoretical physicists as simply just an ordinary interaction
between two objects while any experimentalist or statistician knows that
this is a stochastic process, giving rise to the essential difference
between a \textit{priori\/} and a \textit{posteriori\/} description of the
states.

The last and most essential reason for such an unsatisfactory status of the
quantum measurement theory was the limitations of the projection postulate
applicable only to the instantaneous measurement of the observables with the
discrete spectra, while the real experiments always have a finite duration
and the most important observation is the measurement of the position having
the continuous spectrum.

There are many approaches to the theory of quantum measurement ranging from
purely philosophical to qualitative and even quantitative theories in which
the projection postulate apparently is not needed or is generalized to meet
the indirect, or unsharp, measurements [1--10].\nocite{bib:1,2,3}\nocite%
{bib:4,5,6}\nocite{bib:7,8,9,10}

The most general, the philosophical level, of the discussion of these
problems is of course the simplest and the appropriate one for the largest
audience. But it provides room for unprofessional applications of the more
sophisticated theoretical arguments, giving rise to different kinds of the
speculations and paradoxes. I believe that the professional standard of
quantum measurement theory ought to be an axiomatic and rigorous one and the
quantum measurement problems must be formulated within it and solved
properly instead of making speculations.

In order to examine the quantum paradoxes of Zeno type related to the
continuous measurements, the study must be based on advanced mathematical
methods of the quantum theory of compound systems with not regular but
rather singular interaction, and this has recently received a stochastic
treatment in the quantum theory of open systems and noise. It must use the
tools of the quantum algebraic theory for the calculus of input fields of
the apparatus, i.e., the quantum noises which usually have an infinite
number of degrees of freedom, and for the superselection of output fields,
i.e., commutative (classical) pointer processes which are usually the
stochastic processes in continuous time.

Perhaps some philosophers and physicists would not like such a treatment of
quantum measurement theory; the more mathematical a theory is the less
philosophical it is, and the more rigorous it is, the less alive it is. But
this is just an objective process of the development of any scientific
theory and has already happened with the classical information and
measurement theory.

The corresponding classical dynamical measurement theory, called the
stochastic filtering theory, was developed in the beginning of the 60's by
Stratonovich [11] and for the particular linear case by Kalman [12]. This
theory, based on the notion of the partial (unsharp) observation and the
stochastic calculus method, is optional for the classical deterministic
physics, having dealt with the complete (sharp) observations of the phase
trajectories and ordinary differential calculus, and is usually regarded as
a part of the stochastic systems theory or, more precisely, the classical
information and control theory. The main task of the filtering theory is to
derive and solve a stochastic reduction equation for the present posterior
state of the object of incomplete measurement, giving a means to calculate
the conditional probabilities of the future observations with respect to the
results of the past measurements. The corresponding filtering equation
describes, for example, the continuous spontaneous localization of the
classical Brownian particle under an unsharp observation as the result of
the dynamical reduction of the statistical posterior state given by the
classical conditional expectations under the continuous increase of the
interval of the observation. The stochasticity of this nonlinear equation is
generated either by the Wiener process or by the Poisson process, or by
mixture of them, corresponding to the diffusive, counting, or mixed type of
continuous measurement on the fixed output. It can be also written in the
linear form in terms of the classical renormalized state vector (probability
density), and is sometimes called ``the Schr\"odinger equation of the
classical systems theory'' to emphasize its importance and the probabilistic
interpretation.

Recently the corresponding quantum filtering theory was developed for the
different types of continuous observations, [13,14], although the particular
linear case of quantum Kalman filter was proposed by the author much earlier
[6,7]. This gives rise to an axiomatic quantum measurement theory based on
the new quantum calculus method to handle rigorously the singular
interactions of the quantum object and input fields, and based on the
generalized nondemolition principle to select properly the output observable
processes. The mathematical quantum measurement theory plays the same
central role in the general quantum theory of compound systems containing
the information and control channels. as in the classical systems theory.
But in distinction to the classical case it is not optional for the quantum
physics due to the irreducible probabilistic nature of quantum mechanics
which results in the absence of the phase trajectories. There is no need in
this theory to use the projection or any other reduction postulate. But it
does not contradict the quantum theory, as claimed in Ref. [15], and its
application can be derived in the relevant cases simply as the result of
state vector filtering by means of which the conditional probabilities of
the future observation with respect to the results of the past measurements
are calculated.

There is no need to postulate different nonlinear stochastic modifications
of the Schr\"{o}dinger equation in the phenomenological theories of
spontaneous localization or of the nonstandard quantum theories of dynamical
reduction and continuous collapse, [16--20] and to argue which type is more
universal. They all are given as particular cases [21--24] of the general
diffusive type quantum filtering equation, [25], rigorously derived by
conditioning the corresponding Schr\"{o}dinger equation for the uniquely
determined minimal compound quantum system in Fock--Hilbert space.

The quantum filtering theory gives also a new type of phenomenological
stochastic equations which are relevant to the quantum mechanics with
spontaneous localization, [19,20], corresponding to the random quantum
jumps, [26--28]. This pure discontinuous type is also rigorously derived
from the Schr\"{o}dinger equation [29] by conditioning the
continuous-in-time counting measurement which contains the diffusive type as
the central limit case [30].

Thus, the stochastic nature of measurement processes is reconciled with
unitarity and deterministic interaction on the level of the compound system.
But to account for the unavoidable noise in the continuous observation the
unitary model necessarily involves a quantum system with infinitely many
degrees of freedom and a singular interaction.

The purpose of this paper is to describe explicitly a new universal
nondemolition principle for quantum measurement theory which makes possible
the derivations of the reduction postulates from the quantum interactions.
We show on simple examples what it means to derive rigorously the quantum
filtering equation (thus the Hilbert stochastic process) by conditioning a
Schr\"{o}dinger equation for a compound system. Here, we demonstrate these
derivations from the corresponding unitary interactions with the apparatus
for the particular cases of the measurement of a single observable with the
trivial Hamiltonian $H=0$ of the object using the operator quantum calculus
method instead of the quantum stochastic one [21--23]. But if one wants to
obtain such results in nontrivial cases related to the dynamical observables
that are continuous in time and continuous in spectra and that do not
commute with $H\neq 0$, one needs to use the appropriate mathematical tools,
such as quantum differential calculus and quantum conditional expectations,
recently developed within the algebraic approach in quantum probability
theory. Otherwise, one would be in the same situation as trying to study the
Newton mechanics in nontrivial cases without using the ordinary differential
calculus.

Note that the quantum filtering equation was first obtained in a global form
[9] and then in the differential form [30] within the operational approach,
[1,2], giving the reduced description of the open quantum systems and
quantum continuous measurements. This was done by the stochastic
representation of the continuous instrument, described by the semigroup of
the operational valued measures which are absolutely continuous with respect
to the standard Wiener or Poisson process. The most general approach [31] to
these problems is based on the quantum stochastic calculus of nondemolition
measurements and quantum conditional expectations. It clearly shows that the
operational semigroup approach is restricted to only the Markovian case of
the quantum stochastic object as an open system and to the conditionally
independent nondemolition observations describing the output of the compound
system.

%
%

\section{Causality and nondemolition principle}

Let us begin with the discussion of the quantum nondemolition principle
which forms the basis of the axiomatic formulation of the quantum
measurement theory without the projection postulate, and which has been
implicitly explored also in other approaches [1--10]. The term
``nondemolition measurement'' was first introduced into the theory of
ultrasensitive gravitational experiments by Braginski and others [32--34] to
describe the sequential observations in a quantum Weber antenna as a
simultaneous measurement of some quantum observables. But the property of
nondemolition has never been formalized or even carefully described other
than by requiring the commutativity of the sequential observables in the
Heisenberg picture, which simply means that the measurement process can be
represented as a classical stochastic one by the Gelfand transformation.
Therefore no essentially quantum, noncommutative results have been obtained,
and no theorems showing the existence of such measurements in nontrivial
time continuous models have been proved.

An operator $X$ in a Hilbert space $\mathcal{H}$ is said to be demolished by
an observable $Y=Y^{\dagger }$ in $\mathcal{H}$ if the expectation $\langle
X\rangle $ is changed for $\langle \tilde{X}\rangle \neq \langle X\rangle $
in an initial state when $Y$ has been measured, although without reading.
According to the projection postulate the demolished observable $\tilde{X}%
=\delta \lbrack X]$ is described by the reduction operation $\delta \lbrack
X]=\sum P_{i}XP_{i}$ for a discrete observable $Y=\sum y_{i}P_{i}$ given by
the orthoprojectors $P_{i}^{2}=P_{i}=P_{i}^{\dagger }$, $\sum P_{i}=I$ and
eigenvalues $\{y_{i}\}$. The observable $Y$ is nondemolition with respect to 
$X$ if $\delta \lbrack X]$ is compatible, $\langle \delta \lbrack X]\rangle
=\langle X\rangle $, with respect to each initial state, i.e., iff $\delta
\lbrack X]=X$. It follows immediately in this discrete case that the
nondemolition condition is $XY=YX$, as the main filtering theorem says [30]
even in the general case. Moreover, for each demolition observable $Y$ there
exists a nondemolition representation $\tilde{Y}=\varrho \lbrack Y]$ in an
extended Hilbert space $\mathcal{H}\otimes \mathcal{F}$, which is
statistically equivalent to $Y$ in the sense that $\langle \tilde{X}Y\rangle
=\langle X\tilde{Y}\rangle $ for each input state in $\mathcal{H}$ and
corresponding output state in $\mathcal{H}\otimes \mathcal{F}$. This follows
from the reconstruction theorem [35] for quantum measurements giving the
existence of the nondemolition representation for any kind of observations,
which might be even continuously distributed in the relativistic space-times 
$\mathbb{R}^{1+d}$. In the case of a single discrete observable $Y$ it
proves the unitary reconstruction of the projection postulate, which is
given in section 3.

Now we give several equivalent formulations of the dynamical nondemolition
considered not just as a \emph{possible} property for the quantum
measurements but rather as the \emph{universal} condition to handle such
problems as the modeling of the unsharp measurements, the generalized
reduction and instantaneous collapse for the continuous spectrum
observables, the quantum sequential measurements, and the dynamical
reduction and spontaneous localization under the continuous-in-time
observation. This condition, based on the reconstruction theorem, was
discovered in Ref. [7] and consists of a new principle of quantum axiomatic
measurement theory for the proper representation of the observable process
in a Hilbert space, such as the interaction representation of the object
with the measurement apparatus.

On the philosophical level, one can say that the nondemolition principle is
equivalent to the quantum causality principle of the statistical
predictability for the present and all possible future observations and for
all possible initial states from the a posteriori probability distributions
which are conditioned by the results of the past measurements. This should
be regarded rather as the physical content and purpose of this principle and
not as a definition.

On the mathematical level the nondemolition principle must be formulated as
a necessary and sufficient condition for the existence of the conditional
expectations on the algebras generated by the present and future Heisenberg
operators of the object of the measurement and all the output observables
with respect to the subalgebras of the past measurements and arbitrary input
states.

In the most general algebraic approach this formulation was first obtained
in Ref. [7], (see also Refs. [13] and [14]) as the condition 
\begin{equation}
[X(t),Y(s)]:=X(t)Y(s)-Y(s)X(t)=0\ ,\qquad \forall s\geq t  \label{eq:1.1}
\end{equation}
of compatibility of all system operators $X(t)$ considered as the possible
observables at a time instant $t$ with all past observables $Y(s)$, $s\leq t$%
, which have been measured up to $t$. It says that the Heisenberg operators $%
X(t)$ of the quantum object of the measurement given, say, in the
interaction representation with the apparatus must commute with all past
output observables $Y(s)$, $s\leq t$, of the pointer for any instant $t$.
And according to the causality principle there is no restriction on the
choice of the future observables $Y(r)$, $r\geq t$, with respect to the
present operators $X(t)$ except the self-nondemolition $[Y(r),Y(s)]=0$ for
the compatibility of the family $\{Y(t)\}$. Generalized then in \nocite%
{bib:21,22,23}\nocite{bib:24,25,26}\nocite{bib:27,28} [21--28] for arbitrary 
$X$ and $Y$, these conditions define a stochastic process $Y(t)$ which is
nondemolition with respect to a given quantum process $X(t)$. Note that the
condition (\ref{eq:1.1}) for clearly distinguished object and pointer
observables does not reduce completely the algebra of the compound system to
the commutative one as it does in the case of the direct observations $Y=X$
when it reads as the self-nondemolition condition $[X(t),X(s)]=0$, $\forall
t, s$. The nondemolition measurements considered in Refs. [32--34] were
defined only by the self-nondemolition condition, corresponding to this
trivial (Abelian) case $X(t)=Y(t)$.

In the operational approach [1,2], applicable for the reduced description of
the quantum Markov open system, one might prefer to have a condition that is
equivalent to the nondemolition principle in that case. It can be given in
terms of the induced states on the reduced algebra, i.e., of the states
given by the expectations $\phi (Z)=\langle \psi ,Z(t)\psi \rangle $ on the
algebra of observables $Z$ generated in the Heisenberg picture $%
Z(t)=U^{\dagger }(t)ZU(t)$ by all $X(t)$ and $Y(t)$ for a given initial
state vector $\psi $. The nondemolition principle simply means that the
induced current quantum state of the object coincides with the \textit{a
priori} one, as a statistical mixture of a posteriori states with respect to
the past, but not the future, observations [30]. The a posteriori state as a
quantum state of the object after the measurement, when a result has been
read mathematically, will be defined in the next section. Here we only point
out that the coincidence means that the induced state is not demolished by
the measurement if the results have not been read. This justifies the use of
the word nondemolition in the generalized sense.

One can call this coincidence the generalized reduction principle because it
does not restrict the consideration to the projection valued operations
only, corresponding to the von Neumann reduction of the quantum states,
which is not applicable even for the relatively simple case of instantaneous
measurements of the quantum observables with the continuous spectrum.

The equivalence of these two formulations in the quantum Markovian case and
their relation to the projection postulate (see the next section) can be
illustrated even in the case of the single operation corresponding to an
instantaneous measurement, or a measurement with fixed duration.

Let $\mathcal{H}$ and $\mathcal{F}$ be the Hilbert spaces of state vectors $%
\eta \in \mathcal{H}$, and $\varphi \in \mathcal{F}$ for the quantum object
and the measurement apparatus, respectively, and let $R$ be a self-adjoint
operator in $\mathcal{H}$, representing a dynamical variable with the
spectral values $x\in \mathbb{R}$ to be measured by means of the measurement
apparatus with a given observable $\hat{y}$, representing the pointer of the
apparatus as a self-adjoint operator in $\mathcal{F}$ with either discrete
or continuous spectrum $\Lambda \subseteq \mathbb{R}$. The measurement
apparatus has the fixed initial state $\varphi _{0}\in \mathcal{F}$, $\Vert
\varphi _{0}\Vert =1$ and is coupled to the object by an interaction
operator $S^{\dagger }=V_{0}U^{\dagger }V_{1}$, where $U$ is a unitary
evolution operator of the system in the product space $\mathcal{G}=\mathcal{H%
}\otimes \mathcal{F}$, $U^{\dagger }=U^{-1}$, and $V_{0}=V\otimes \hat{1}$, $%
V_{1}=I\otimes \hat{v}$ are the unitarities given by the free evolution
operators $V:\mathcal{H}\rightarrow \mathcal{H}$, $\hat{v}:\mathcal{F}%
\rightarrow \mathcal{F}$ of the object and the apparatus, respectively,
during the fixed measurement interval $[0,t]$. It is natural to suppose that
the interaction does not disturb the variable $R$ in the sense $%
R_{0}:=R\otimes \hat{1}=S^{\dagger }R_{0}S$, or equivalently, $\langle x|S=%
\hat{s}_{x}\langle x|$, i.e., 
\begin{equation}
S:|x\rangle \otimes \varphi _{0}\mapsto |x\rangle \otimes \varphi _{x}\
,\quad \forall x\in \mathbb{R}  \label{eq:1.2}
\end{equation}%
in terms of (generalized) eigenvectors $|x\rangle $ of $R$, where $\varphi
_{x}=\hat{s}_{x}\varphi _{0}$. But it must disturb the input observable $%
\hat{q}=\hat{v}^{\dagger }\hat{y}\hat{v}$ in order to get the
distinguishable probability densities $f_{x}(y)=|\varphi _{x}(y)|^{2}$ of
the output observable $Y=S^{\dagger }(\kappa I\otimes \hat{q})S$,
corresponding to the different eigen values $x\in \mathbb{R}$ of the input
states $|x\rangle $ to be tested by the usual methods of mathematical
statistics. Here $\kappa >0$ is a scaling parameter and we have assumed, for
simplicity that the observable $\hat{y}$ and hence $\hat{q}$ has the
nondegenerate spectral values $y\in \Lambda $, so that $\varphi \in \mathcal{%
F}$ in the input representation is described by the (generalized)
eigenvectors $|y\rangle $ of $\hat{q}:|y\rangle \mapsto y|y\rangle $ as a
square integrable function $\varphi (y)=\langle y|\varphi $, $\Vert \varphi
\Vert ^{2}=\int |\varphi (y)|^{2}\mathrm{d}\nu <\infty $ with respect to a
given measure $\nu $ on $\Lambda $.

The positive measure $\nu $ is either discrete or continuous or can even be
of mixed type normalizing the probability densities $g(y)=\langle \psi
(y),\psi (y)\rangle $ for the state vectors $\psi \in \mathcal{G}$: 
\begin{equation}
\Vert \psi \Vert ^{2}=\int_{\Lambda }\langle \psi (y),\ \psi (y)\rangle 
\mathrm{d}\nu =\int_{\Lambda }g(y)\mathrm{d}\nu =1  \label{eq:1.4}
\end{equation}%
where $\psi (y)=\langle y|\psi $ are the $\mathcal{H}$-valued wavefunctions
of the system \textquotedblleft quantum object plus measurement
apparatus.\textquotedblright\ One can consider, for example, the standard
Lebesgue measures $\mathrm{d}\nu =\mathrm{d}\lambda $ on $\Lambda =\mathbf{Z}
$, $\mathrm{d}\lambda =1$ and on $\Lambda =\mathbb{R}$, $\mathrm{d}\lambda =%
\mathrm{d}y$: 
\begin{equation*}
\Vert \psi \Vert ^{2}=\sum \langle \psi (k),\psi (k)\rangle \;(\mathrm{d}%
\lambda =1)\ ;\quad \Vert \psi \Vert ^{2}=\int \langle \psi (y),\psi
(y)\rangle \mathrm{d}y\;(\mathrm{d}\lambda =\mathrm{d}y) 
\end{equation*}%
respectively for the discrete spectrum $y\in \mathbf{Z}$ and for the
continuous one $y\in \mathbb{R}$, given by the distributions $f(y)=\sum
\delta (y-k)$ and $f(y)=1$ as $\mathrm{d}\lambda =f(y)\mathrm{d}y$.

The output state vectors $\chi =S(\xi \otimes \varphi _{0})\in \mathcal{G}$,
corresponding to the arbitrary input ones $\xi =V\eta $, $\langle \xi ,\xi
\rangle =1$, are given by the vector-functions $\chi :y\mapsto \chi (y)\in 
\mathcal{H}$ of $y\in \Lambda $ with values%
\begin{equation*}
\chi (y)=\langle y|S(\xi \otimes \varphi _{0})=\langle y|\chi .
\end{equation*}%
The operators $\langle y|S:\mathcal{G}\rightarrow \mathcal{H}$ correspond to
the adjoint ones $S^{\dagger }|y\rangle :\eta \mapsto S^{\dagger }(\eta
\otimes |y\rangle )$, 
\begin{equation}
\langle \eta ,\langle y|S(\xi \otimes \varphi )\rangle =\langle S^{\dagger
}(\eta \otimes |y\rangle ),\xi \otimes \varphi \rangle   \label{eq:1.3}
\end{equation}%
defining the (generalized) vector-functions $S^{\dagger }|y\rangle \eta $ by 
\begin{equation*}
\int S^{\dagger }|y\rangle \eta \varphi _{0}(y)\mathrm{d}\nu =S^{\dagger
}(\eta \otimes \varphi _{0})\;\;\;\;\forall \eta ,\varphi .
\end{equation*}%
The operator $(R_{0}\chi )(y)=R\chi (y)$ commutes with $Q=\kappa I\otimes 
\hat{q}$ as well as with any other operator $C_{0}=C\otimes \hat{1}$
representing an object variable $C:\mathcal{H}\rightarrow \mathcal{H}$ in $%
\mathcal{H}\otimes \mathcal{F}$ as the constant function $Z(y)=C$. This is
because the general operator $Z$ in $\mathcal{H}\otimes \mathcal{F}$
commuting with $Q$ corresponds to an operator--valued function $Z(y):%
\mathcal{H}\rightarrow \mathcal{H}$, which is defined by the operator $Z$ as 
\begin{equation}
\langle y|Z\psi =Z(y)\langle y|\psi \ ,\quad \forall \psi \in \mathcal{G}\
,\quad y\in \Lambda   \label{eq:1.5}
\end{equation}%
in the case $Z=Q$ it corresponds to $Z(y)=\kappa yI$: $\ \langle y|Q\psi
=\kappa y\langle y|\psi $. It is trivial in this case that the Heisenberg
operators $X=S^{\dagger }ZS$ satisfy the nondemolition condition $[X,Y]=0$
with respect to the output observable $Y=S^{\dagger }QS$, but not the
initial operators $Z:[Z,Y]\not=0$ if $[Z(y),R]\not=0$. This makes it
possible to condition, by the observation of $Y$, the future measurements of
any dynamical variable of the quantum object, but not the potential
measurements of $Z$ in the past with respect to the present observation of $Y
$ if they have not been done initially.

Indeed, let $P_{\Delta }=S^{\dagger }I_{\Delta }S$ be the spectral
orthoprojector of $Y$, given for a measurable $\Delta \subseteq \Lambda $ by 
$I_{\Delta }=I\otimes \hat{1}_{\Delta }$ as 
\begin{equation}
\langle y|I_{\Delta }\chi =1_{\Delta }(y)\chi (y)=1_{\Delta }(y)\langle
y|\chi \ ,\quad 1_{\Delta }(y)=\{%
\begin{array}{cc}
1 & y\in \Delta \\ 
0 & y\neq \Delta%
\end{array}
\label{eq:1.6}
\end{equation}%
and $p_{\Delta }=\langle \eta \otimes \varphi ,P_{\Delta }(\eta \otimes
\varphi )\rangle \not=0$. Then the formula 
\begin{equation}
\varepsilon _{\Delta }[X]=\langle \eta ,\omega \lbrack XP_{\Delta }]\eta
\rangle /\langle \eta ,\omega \lbrack P_{\Delta }]\eta \rangle \,
\label{eq:1.7}
\end{equation}%
where $\langle \eta ,\omega \lbrack X]\eta \rangle =\langle \eta \otimes
\varphi ,X(\eta \otimes \varphi )\rangle $, $\forall \eta \in \mathcal{H}$,
defines the conditional expectation of $X=S^{\dagger }ZS$ with respect to $Y$%
. It gives the conditional probability $\varepsilon _{\Delta }[X]\in \lbrack
0,1]$ for any orthoprojector $X=O$, while $\varepsilon _{\Delta }[Z]$
defined by the same formula for $Z=\{Z(y)\}$ may not be the conditional
expectation due to the lack of positivity $\omega \lbrack EP_{\Delta }]\geq
0 $, for all $\varphi \in \mathcal{F}$ if the orthoprojector$Z=E$ does not
commute with $P_{\Delta }$. The necessity of the nondemolition principle for
the existence of the conditional probabilities is the consequence of the
main filtering theorem consistent with the causality principle according to
which the conditioning with respect to the current observation has the sense
of preparation for future measurements but not for past ones.

This theorem proved in the general algebraic form in Ref. [30] reads in the
simplest formulation as\medskip

\noindent \textsc{Main Measurement Theorem.} Let $O$ be an orthoprojector in 
$\mathcal{G}=\mathcal{H}\otimes \mathcal{F}$. Then for each state vector $%
\psi =\xi \otimes \varphi $ there exists the conditional probability $%
\varepsilon _{\Delta }[O]\in \lbrack 0,1]$, defined by the compatibility
condition 
\begin{equation}
\varepsilon _{\Delta }[O]\langle \xi \otimes \varphi ,P_{\Delta }(\xi
\otimes \varphi )\rangle =\langle \xi \otimes \varphi ,\ OP_{\Delta }(\xi
\otimes \varphi )\rangle  \label{eq:1.8}
\end{equation}%
if and only if $OP_{\Delta }=P_{\Delta }O$. It is uniquely defined for any
measurable $\Delta \subset \Lambda $ with respect to $P_{\Delta }=S^{\dagger
}I_{\Delta }S$, $\varphi =\varphi _{0}$ as 
\begin{equation}
\varepsilon _{\Delta }[O]={\frac{1}{\mu _{\Delta }}}\int_{\Delta }\langle
\chi _{y},E(y)\chi _{y}\rangle \mathrm{d}\mu  \label{eq:1.9}
\end{equation}%
Here $E(y):\mathcal{H}\rightarrow \mathcal{H}$ is the orthoprojector valued
function, describing $O$, commuting with all $P_{\Delta }$ in the Schr\"{o}%
dinger picture as $O=S^{\dagger }ES$, $\mu _{\Delta }=\int_{\Delta }g_{\xi
}(y)\mathrm{d}\nu $ is the absolutely continuous with respect to $\nu $
probability distribution of $y\in \Lambda $, $g_{\xi }(y)=\Vert \chi
(y)\Vert ^{2}$, $\chi (y)=\langle y|S(\xi \otimes \varphi _{0})$, and $%
y\mapsto \chi _{y}$ is the random state vector $\chi _{y}\in \mathcal{H}$ of
the object uniquely defined for almost all $y:g_{\xi }(y)\not=0$ up to the
random phase $\theta (y)=\mathrm{arg}c_{\xi }(y)$ by the normalization 
\begin{equation}
\chi _{y}=\chi (y)/c_{\xi }(y)\ ,\quad |c_{\xi }(y)|^{2}=g_{\xi }(y)
\label{eq:1.10}
\end{equation}

\section{The generalized \emph{a posteriori} reduction}

It follows immediately from the main theorem that the input state vector $%
\xi :\Vert \xi \Vert =1$ of the object of measurement has to be changed for $%
\chi _{y}\in \mathcal{H}$ due to the preparation $\xi \mapsto \{\chi
(y):y\in \Lambda \}$ of the \textit{a priori\/} state vector $\chi =S(\eta
\otimes \varphi _{0})$ of the meter and the object after the objectification 
$\hat{q}=y$. The former is given by the dynamical interaction in the pointer
representation $\chi (y)=\langle y\mid \chi $ due to the choice of the
measurement apparatus and the output observables, and the latter is caused
by statistical filtering $\chi \mapsto \chi (y)$ due to the registration of
the measurement result $y\in \Lambda $ and the normalization $\chi _{y}=\chi
(y)/\Vert \chi (y)\Vert $.

While the process of preparation described by a unitary operator applied to
a fixed initial state of the meter encounters no objection among physicists,
the process of objectification encounters objection because of the
nonunitarity of the filtering and nonlinearity of the normalization. But the
main theorem shows clearly that there is nothing mysterious in the
objectification. It is not a physical process but only a mathematical
operation to evaluate the \textit{conditional state} 
\begin{equation}
\pi _{y}[Z]=\varepsilon _{y}[S^{\dagger }ZS]=\langle \chi _{y},Z(y){\chi _{y}%
}\rangle   \label{eq:2.1}
\end{equation}%
which are defined by the conditional expectations $\varepsilon
_{y}[X]=\lim_{\Delta \downarrow y}\varepsilon _{\Delta }[X]$ of the
Heisenberg operators $X$ for $Z=\{Z(y)\}$. The linear random operator 
\begin{equation}
G(y):\xi \in {\mathcal{H}}\mapsto \langle y|S(\xi \otimes \varphi _{0})\
,\quad y\in \Lambda   \label{eq:2.2}
\end{equation}%
defines the reduction transformations $G(y)$ as the partial matrix elements $%
\langle y|S\varphi _{0}$ of the unitary operator $S$. They map the
normalized vectors $\xi \in \mathcal{H}$ into the \emph{a posteriori\/} ones 
$\chi (y)=G(y)\xi $, renormalized to the probability density 
\begin{equation*}
g_{\xi }(y)=\Vert G(y)\xi \Vert ^{2}=\langle \xi ,E(y)\xi \rangle ,\quad
E=G^{\dagger }G\ .
\end{equation*}%
If the condition (\ref{eq:1.2}) holds, then the only eigen vectors $%
|x\rangle $ of $R$ remain unchanged up to a phase multiplier: 
\begin{equation}
G(y)|x\rangle =|x\rangle \varphi _{x}(y),\ \varphi _{x}(y)=\langle y|\hat{s}%
_{x}\varphi _{0}=\langle y|\varphi _{x}  \label{eq:2.3}
\end{equation}%
and hence $\chi _{y}=e^{\mathrm{i}\theta _{x}(y)}|x\rangle $, where $\theta
_{x}(y)=\arg \,\varphi _{x}(y)$. The superpositions $\xi =\int |x\rangle \xi
(x)\mathrm{d}\lambda $ change their amplitudes $\xi (x)=\langle x|\xi $ for $%
\chi _{y}(x)=\langle x|\chi _{y}$ 
\begin{equation}
\langle x|\chi _{y}=c_{\xi }^{-1}(y)\chi (x,y)\ ,\quad \chi (x,y)=\langle
x|G(y)\xi =\chi _{x}(y)\xi (x)  \label{eq:2.4}
\end{equation}%
where $c_{\xi }(y)=(\int |\varphi _{x}(y)|^{2}h(x)\mathrm{d}\lambda )^{1/2}$%
, $h(x)=|\xi (x)|^{2}$.

In the case of a purely continuous spectrum of $R$ there are no invariant
state vectors at all because the generalized eigenvectors cannot be
considered as input ones due to $|x\rangle \notin \mathcal{H}$ as $\langle
x|x\rangle=\infty$ in that case.

The generalized reduction (\ref{eq:2.1}) of the state-vector corresponds to
the limit case $\Delta\downarrow y$ when the accuracy of the instrument $%
\Delta\ni y$ tends to the single-point subset $\{y\}\subset\Lambda$. It is
not even the mathematical idealization of the real physical experiment if
the observable $\hat q$ has the discrete spectrum $\Lambda=\{y_i\}$.

Prior to discussing why the generalized reduction does not contradict the
main postulates of the quantum theory, let us show how to derive the von
Neumann projection postulate in this way, corresponding to the orthogonal
transformations $G(y_{i})=F_{i}$ given by a partition $\sum A_{i}=\mathbb{R}$
of the spectrum of $R$ as $F_{i}=E_{A_{i}}$. Here $A\mapsto E_{A}$, $%
E_{A}^{\dagger }E_{A^{\prime }}=E_{A\cap A^{\prime }}$, $\sum E_{A_{i}}=I$
is the spectral measure of $R=\int x\mathrm{d}E$ which might be either of
discrete or of continuous type as in the cases 
\begin{equation*}
E_{A}=\sum_{x\in A}|x\rangle \langle x|\ ,\quad E_{A}=\int_{A}|x\rangle
\langle x|\mathrm{d}x\ , 
\end{equation*}%
corresponding to the nondegenerate spectrum of $R:\mathrm{d}E=|x\rangle
\langle x|\mathrm{d}\lambda $.

Considering the indices $i$ of $y_{i}$ in $\mathbf{Z}$ it is always possible
to find the unitary interaction in the Hilbert space ${\mathcal{G}=\mathcal{H%
}}\otimes l^{2}(\mathbf{Z})$ of the two--sided sequences $\psi =\{\eta
^{k}|k=0,\pm 1,\pm 2,\dots \}$ with $\eta ^{k}\in \mathcal{H}$ such that $%
\Vert \psi \Vert ^{2}=\sum_{-\infty }^{\infty }\langle \eta ^{k},\eta
^{k}\rangle <\infty $. Indeed, we can define the interaction as the
block-matrix $S^{\dagger }=[W_{k}^{i}]$ acting in $\mathcal{G}$ as $%
W^{i}\psi =\sum_{k=-\infty }^{\infty }W_{k}^{i}\eta ^{k}$, by $%
W_{k}^{i}=F_{k-i}$, where $F_{k}=0$ if there is no point $y_{k}$ in $\Lambda 
$ numbered by a $k\in \mathbf{Z}$. It is the unitary one because $S=[F_{i-k}]
$ is inverse to $S^{\dagger }=[F_{k-i}]$ as 
\begin{equation*}
\sum_{j=-\infty }^{\infty }F_{i-j}F_{k-j}=\delta _{k-j}^{i-j}\sum_{j=-\infty
}^{\infty }F_{-j}=\delta _{k}^{i}\sum F_{i}=\delta _{i}^{k}I 
\end{equation*}%
due to the orthogonality $F_{i}F_{k}=0$, $i\not=k$, and completeness $\sum
F_{i}=I$ of $\{F_{i}\}$.

Let us fix the initial sequence $\varphi _{0}\in l^{2}(\mathbf{Z})$ as the
eigenstate $\varphi _{0}=\{\delta _{0}^{k}\}=|0\rangle $ of the input
observable $\hat{k}$ in $l^{2}(\mathbf{Z})$ as the counting operator 
\begin{equation}
\hat{k}=\sum_{k=-\infty }^{\infty }k|k\rangle \langle k|\ ,\quad |i\rangle
=\{\delta _{i}^{k}\}\in l^{2}(\mathbf{Z})  \label{eq:2.5}
\end{equation}%
with the spectrum $\mathbf{Z}$. Then we obtain the conditional states (\ref%
{eq:2.1}) defined as 
\begin{equation*}
\pi _{i}[Z]={\frac{1}{p_{i}}}\langle F_{i}\eta ,Z_{i}F_{i}\eta \rangle
=\langle \eta _{i},Z_{i}\eta _{i}\rangle ,\;\eta _{i}=F_{i}\eta /p_{i}^{1/2} 
\end{equation*}%
up to the normalizations $p_{i}=\langle F_{i}\eta ,F_{i}\eta \rangle \not=0$
by the linear operations $\sigma \mapsto W_{i}^{0}\sigma W_{i}^{0}$, 
\begin{equation}
W_{i}^{0}\eta =\langle i|S(\varphi _{0}\otimes \eta )=\sum_{k=-\infty
}^{\infty }F_{i-k}\delta _{0}^{k}\eta =F_{i}\eta \ .  \label{eq:2.6}
\end{equation}%
It is only in that case that the \emph{a posteriori\/} state always remains
unchanged under the repetitions of the measurement. Such an interaction
satisfies the condition (\ref{eq:1.2}) with $\varphi _{x}=\hat{s}_{x}\varphi
_{0}$ given by the sequences $\varphi _{x}=\{\delta
_{i(x)}^{k}\}=|i(x)\rangle $ because 
\begin{equation*}
F_{i-k}|x\rangle =|x\rangle \delta _{i-k}^{i(x)}=W_{i}^{k}|x\rangle \quad
(=|x\rangle \,,\quad \forall x\in A_{i-k})\,, 
\end{equation*}%
where $i(x)=i$ if $x\in A_{i}$ is the index map of the coarse-graining $%
\{A_{i}\}$ of the spectrum of $R$. Hence in the $x$-representation $\psi
=\int |x\rangle \psi (x)\mathrm{d}\lambda $, $\psi (x)=\langle x|\psi $ it
can be described by the shifts $\hat{s}_{x}^{\dagger }=[\delta _{k-i}^{i(x)}]
$ in $l^{2}(\mathbf{Z})$ 
\begin{equation}
\hat{s}_{x}^{\dagger }:\psi (x)=\{\eta ^{k}(x)\}\mapsto \{\langle x|\eta
^{i(x)+k}\}\,\quad \eta ^{k}(x)=\langle x|\eta ^{k}  \label{eq:2.7}
\end{equation}%
replacing the initial state $\varphi _{0}=|0\rangle $ of the meter for each $%
x$ by another eigenstate $|i(x)\rangle =\hat{s}_{x}|0\rangle $ if $x\notin
A_{0}$.

This realizes the coarse-grained measurement of $R$ by means of the
nondemolition observation of the output 
\begin{equation}
Y=S^{\dagger }(I\otimes \hat{k})S=i(R)\otimes \hat{1}+I\otimes \hat{k}\,,
\label{eq:2.8}
\end{equation}%
where $i(R)=\int i(x)\mathrm{d}E=\sum iF_{i}$. If $q(R)=\hbar i(R)$ is the
quantized operator $R$ given, say, by the integer $i(x)=\lfloor x/\hbar
\rfloor $, then the rescaled model $\hat{y}_{x}=\hbar \hat{s}_{x}^{\dagger }%
\hat{k}\hat{s}_{x}=q(x)\hat{1}+\hbar \hat{k}$ of the nondemolition
measurement has the classical limit $\lim \hat{y}_{x}=x\hat{1}$ if $\hbar
\rightarrow 0$, corresponding to the direct observation of a continuous
variable $R$ by means of $\lim \hbar Y=R\otimes \hat{1}$.

Note that the observable $Y$ commutes with the arbitrary Heisenberg operator 
$A=S^{\dagger }(C\otimes \hat{1})S$ of the object, but not with the initial
operators $C_{0}=C\otimes \hat{1}$ if $[C,i(R)]\not=0$.

The unitary operator $S^{\dagger }$ is given by the interaction potential $%
q(R)\otimes \hat{p}$ as $S^{\dagger }=\exp \{(\mathrm{i}/\hbar )q(R)\otimes 
\hat{p}\}$, where $\mathrm{i}=\sqrt{-1}$, and $\hat{p}=[\langle i|\hat{p}%
|k\rangle ]$, $\langle i|\hat{p}|k\rangle =(1/2\pi )\int_{-\pi }^{\pi }pe^{-%
\mathrm{i}(i-k)p}\mathrm{d}p$ is the matrix of the momentum operator in $%
l^{2}(\mathbf{Z})$, generating the shifts $\hat{s}_{x}^{\dagger }=[\langle i|%
\hat{s}_{x}^{\dagger }|k\rangle ]$ as $\hat{s}_{x}^{\dagger }=e^{i(x)\mathrm{%
i}\hat{p}} $: 
\begin{equation*}
\langle i|\hat{s}_{x}^{\dagger }|k\rangle ={\frac{1}{2\pi }}\int_{-\pi
}^{\pi }e^{i(x)\mathrm{i}p}e^{-\mathrm{i}(i-k)p}\mathrm{d}p=\delta
_{i-k}^{i(x)}\,. 
\end{equation*}%
The nondemolition observation reproduces the statistics of the
\textquotedblleft demolition\textquotedblright\ measurement of $R$ by the
direct observation of $q(R)$ because the output observable $Y$ has the same
characteristic function with respect to the state vector $\xi \otimes
\varphi _{0}$ as $i(R)$ with respect to $\xi $: 
\begin{eqnarray*}
&\langle \xi \otimes \varphi _{0},\exp \{\mathrm{i}pY\}(\xi \otimes \varphi
_{0})\rangle =&\langle S(\xi \otimes \varphi _{0}),e^{\mathrm{i}pQ}S(\xi
\otimes \varphi _{0})\rangle \\
&&\qquad =\sum \langle F_{i}\xi ,e^{i\mathrm{i}p}F_{i}\xi \rangle =\langle
\xi ,\exp \{\mathrm{i}pi(R)\}\xi \rangle \,.
\end{eqnarray*}%
Here $p$ is the parameter of the characteristic function, $Q=I\otimes \hat{k}
$, and $F_{i}=\langle i|S\varphi _{0}=F_{i}^{\dagger }$ are the
orthoprojectors, such that $\sum_{i}F_{i}^{\dagger }F_{i}=\int i(x)\mathrm{d}%
E=i(R)$. If the observable $R$ is discrete, then the nondemolition
observation (\ref{eq:2.8}) realizes the precise measurement of $R$, if the
partition $\{A_{i}\}$ separates all the eigenvalues $\{x_{i}\}$ as in the
case $x_{i}\in A_{i}$, $\forall i$, corresponding to $x_{i}=\hbar i$, $%
A_{i}=[\hbar i,\hbar (i+1)[$, $i=0,1,2,\ldots $.

The nondemolition principle helps not only to derive the projection
postulate as a reduced description of the shift interaction in the enlarged
Hilbert space $\mathcal{G}$ with respect to the initial eigenvector $\varphi
_{0}=|0\rangle $ of the discrete meter $\hat{q}$, but also extends it to the
generalized reductions under the unsharp measurements with arbitrary
spectrum $\Lambda $, corresponding to the nonrepeatable instruments [1,2] 
\begin{equation}
\Pi _{\Delta }[C]=\int_{\Delta }\Psi \lbrack C](y)\mathrm{d}\nu \,,\quad
\Psi \lbrack C](y)=G(y)^{\dagger }CG(y)\,.  \label{eq:2.9}
\end{equation}%
The density $\Psi (y)$ of the instrument defines completely positive but not
necessarily orthoprojective operations $E(y)=\Psi \lbrack I](y)$, called the
effects for the probability densities $g(y)=\sigma \lbrack E(y)]$, and also
the nonlinear operation $\sigma \mapsto \sigma \circ \Psi (y)/\sigma \lbrack
E(y)]$ of the generalized reduction, mapping the pure input states $\sigma
_{\xi }[C]=\langle \xi |C|\xi \rangle $ into the \emph{a posteriori\/} ones 
\begin{equation}
\rho _{y}[C]={\frac{1}{g_{\xi }(y)}}\rho \lbrack C](y)=\pi _{y}[C_{0}]\
,\quad \rho \lbrack C](y)=\langle \chi (y),\ C\chi (y)\rangle \,.
\label{eq:2.10}
\end{equation}%
They are also pure because of the completeness of the nondemolition
measurement, i.e., nondegeneracy of the spectrum of the observable $\hat{q}$
in $\mathcal{F}$. Thus, the reduction of the state-vector is simply the way
of representing in the form (\ref{eq:2.1}) the \emph{a posteriori\/} pure
states (\ref{eq:2.10}) given at the limit $\Delta \rightarrow 0$ by the
usual (in the statistics) Bayesian formula (\ref{eq:1.7}) for $X=S^{\dagger
}C_{0}S=A$, which is applicable due to the commutativity of $A$ and $%
P_{\Delta }$.

The reduction $\sigma _{1}\rightarrow \rho _{y}$ of the prepared state $%
\sigma _{1}=\sigma \circ \Psi $ for the object measurement is given as the
evaluation of the conditional expectations which are the standard attributes
of any statistical theory. All the attempts to derive the reduction as a
result of deterministic interaction only are essentially the doomed attempts
to derive the probabilistic interpretation of quantum theory. There is no
physical explanation of the stochasticity of the measurement process as
there is no adequate explanation of the randomness of an observable in a
pure quantum state.

It is not a dynamical but a purely statistical effect because the input and
output state-vectors of this process are not the observables of the
individual object of the statistical ensemble but only the means for
calculating the \textit{a priori}\emph{\/} and the \textit{a posteriori}%
\emph{\/} probabilities of the observables of this object. Hence there is no
observation involving just a single quantum object which can confirm the
reduction of its state. The reduction of the state-vector can be treated as
an observable process only for an infinite ensemble of similar object plus
meter systems. But the measurements for the corresponding collective
observables also involves preparation and objectification procedures, this
time for the ensemble, i.e., for a second quantized compound system. So the
desirable treatment of all the reductions as some objective stochastic
process can never be reached in this way. They are secondary stochastic
since they are dependent on the random information that has been gained up
to a given time instant $t$.

The reduction of the state-vector is not at variance with the coherent
superposition principle, because a vector $\eta \in \mathcal{H}$ is not yet
a pure quantum state but defines it rather up to a constant $c\in \mathbb{C}$
as the one-dimensional subspace $\{c\eta |c\in \mathbb{C}\}\subset \mathcal{H%
}$ which is a point of the projective space over $\mathcal{H}$. For every
reduced state-vector $\chi _{y}$ there exists an equivalent one, namely $%
\chi (y)=\sqrt{g_{\xi }(y)}\chi _{y}$, defining the same quantum pure state,
given by the linear transformation $G(y):\xi \mapsto \chi (y)$, so that the
superposition principle holds: $\chi (y)=\sum c_{i}\chi ^{i}(y)$ if $\xi
=\sum c_{i}\xi ^{i}$. The pure state transformation $G(y)$ does not need to
be unitary, but as an operator $G:\mathcal{H}\rightarrow \mathcal{G}$ with 
\begin{equation*}
G^{\dagger }G=\int G(y)^{\dagger }G(y)\mathrm{d}\nu =\int \varphi
_{0}^{\dagger }S^{\dagger }|y\rangle \langle y|S\varphi _{0}\mathrm{d}\nu
=\varphi _{0}^{\dagger }S^{\dagger }S\varphi _{0}=I 
\end{equation*}%
it preserves the total probability by mapping the normalized $\xi \in 
\mathcal{H}$ into the $\chi (y)=G(y)\xi $, normalized to the probability
density $g_{\xi }(y)$.

According to the nondemolition principle it makes sense to apply the vector $%
\chi=\{\chi(y)\}$ of the system after the measurement preparation only
against the reduced observables $Z=\{Z(y)\}$ which commute with $Q=\kappa
I\otimes \hat q$. Otherwise according to the main theorem the conditional
probabilities of the future observations may not exist for an initial
state-vector $\chi_0=\eta\otimes\varphi$ and a given result $y\in\Lambda$ of
the measurement. It is against the physical causality to consider the
unreduced operators as the observables for the future measurements since the
causality means that the future observations must be statistically
predictable from the data of a measurement and such prediction can be given
only by the conditional probabilities (\ref{eq:1.9}). Once the output
observables are selected as a part of a preparation, the algebra of the
actual observables is reduced and there is no way to measure an observable $%
Z $ which is not compatible with $Q$. It could be measured in the past if
another preparation had been made but the irreversibility of the time arrow
does not give this possibility. Thus, the quantum measurement theory implies
a kind of time-dependent superselection rule for algebras such as those of
the observables $Z$ chosen as the actual observable at the moment $t$. But
it does not prevent one from considering other operators as the virtual
observables defining super operators, i.e., the subsidiary operators for the
description of some meaningful operations, although an evaluation of their
expectations does not make any sense as it does for the differential
operators in the classical theory.

The \textit{a priori}\emph{\/} states are the induced ones 
\begin{equation*}
\sigma _{1}(C)=\int \langle \chi _{y},C\chi _{y}\rangle \mathrm{d}\mu
=\langle \chi ,C_{0}\chi \rangle \ ,\quad C_{0}=C\otimes \hat{1}
\end{equation*}%
on the algebra generated by the operators in $\mathcal{H}$ of the object
only. They are given as the statistical mixtures of the \textit{a posteriori}%
\emph{\/} pure states (\ref{eq:2.10}) of the object even if the initial
state $\sigma $ was pure. But it does not contradict quantum mechanics
because the prepared state $\phi (Z)=\langle \chi ,Z\chi \rangle $ of the
quantum system after the measurement is reduced to the object plus pointer
but is still given uniquely by the state-vector $\chi \in \mathcal{G}$, up
to a random phase. Namely, the vector $\chi $ is a coherent superposition 
\begin{equation*}
\chi =\sum \chi _{i}\otimes |y_{i}\rangle c_{i}\ ,\quad \chi _{i}=\chi
(y_{i})/c_{i}\ ,\quad |c_{i}|^{2}=p_{i}
\end{equation*}%
of the \textit{a posteriori}\emph{\/} states $\chi _{i}\otimes |y_{i}\rangle 
$ of the system, if $\hat{q}$ has the spectral decomposition $\hat{q}=\sum
y_{i}|y_{i}\rangle \langle y_{i}|$ and $p_{i}$ are the probabilities of $%
y_{i}$.

This uniqueness does not hold for the density-matrix representations $\phi
\lbrack Z]=\mathrm{Tr}\{\hat{\phi}Z\}$; among the equivalent density
matrices $\hat{\phi}\geq 0$ there exists always the projector $\hat{\phi}%
=|\chi \rangle \langle \chi |$, but there are also mixtures such as the
diagonal one 
\begin{equation*}
\hat{\phi}_{1}=\sum p_{i}|\eta _{i}\rangle \langle \eta _{i}|\ ,\quad |\eta
_{i}\rangle =\eta _{i}\otimes |y_{i}\rangle 
\end{equation*}%
in the discrete case $\Lambda =\{y_{i}\}$. Hence, the diagonalization $\hat{%
\phi}\mapsto \hat{\phi}_{1}$ of the density matrix due to the measurement of 
$\hat{q}$ is only the rule to choose the most mixed one $\hat{\phi}_{1}$
which is equivalent to the coherent choice $\hat{\phi}$ due to 
\begin{equation*}
\mathrm{Tr}\{\hat{\phi}Z\}=\sum p_{i}\langle \eta _{i},Z_{i}\eta _{i}\rangle
=\mathrm{Tr}\{\hat{\phi}_{1}Z\} 
\end{equation*}%
for all reduced operators $Z=\sum Z_{i}\otimes |y_{i}\rangle \langle y_{i}|$%
. There is no special need to fix such a choice, which is even impossible in
the continuous spectrum case. This is because the continuous observable $%
\hat{q}$ has no ordinary eigenvectors, $\langle y|y\rangle =\infty $ and
hence $\chi _{y}\otimes |y\rangle \notin \mathcal{G}$, but there exist the
eigenstates $\omega _{y}[\hat{z}]=z(y)$ on the algebra of complex functions $%
z(y)$, defining the conditional expectations $\varepsilon _{y}[X]$ for $%
X=S^{\dagger }ZS$ as 
\begin{equation*}
\varepsilon _{y}[X]=\pi _{y}[SXS^{\dagger }]\ ,\quad \pi _{y}=\rho
_{y}\otimes \omega _{y}\ ,\quad \forall y\in \Lambda \,. 
\end{equation*}%
Thus, the nondemolition principle abandons the collapse problem, reducing it
to the evaluation of the \emph{a posteriori\/} state. The decrease of the
observable algebra is the only reason for the irreversibility of the linear
transformation $\phi _{0}\mapsto \phi $ of the initial states $\phi
_{0}(X)=\langle \chi _{0},X\chi _{0}\rangle $, which are pure on the algebra
of all operators $X$ into the prepared (mixed) ones on the algebra of the
reduced operators $Z$.

\section{The main measurement problem}

As was shown using an instantaneous measurement as an example, the
nondemolition principle leads to the notion of the instrument, described by
the operational-valued measure (\ref{eq:2.9}), and gives rise to the
generalized reduction (\ref{eq:2.10}) of the quantum statistical states. In
the operational approach [1,2] one starts from the instrumental description $%
\sigma\mapsto\sigma\circ\Phi(y)=\rho(y)$ of the measurement, which is
equivalent to postulating the generalized reduction (20) given up to the
probabilistic normalization $g(y)=\rho[I](y)$ by the linear map $%
\sigma\mapsto\sigma\circ\Psi(y)$ due to $\Psi_y(\sigma)=(1/g(y))\sigma\circ%
\Psi(y)=\rho_ y$.

The main measurement problem is the reconstruction of an interaction
representation of the quantum measurement, that is, finding a proper
dilation $\mathcal{G}$ of the Hilbert space $\mathcal{H}$ and the output
process $Y $, satisfying the nondemolition (and self-nondemolition)
condition (\ref{eq:1.1}) with respect to the Heisenberg operators $X$ of the
object of measurement in order to derive the same reduction as the result of
conditional expectation.

The minimal dilation giving, in principle, the solution of this problem even
for non-Markovian relativistic cases was constructed in [35], but it is
worth finding also more realistic, nonminimal dilations defining the object
of measurement as a quantum stochastic process in the strong sense for the
particular Markovian cases.

In the case of a single instantaneous measurement described by an instrument 
$\Pi _{\Delta }$, this can be formulated as the problem of finding the
unitary dilation $U\varphi _{0}:\eta \in \mathcal{H}\mapsto U(\eta \otimes
\varphi _{0})$ in a tensor product $\mathcal{G}=\mathcal{H}\otimes \mathcal{F%
}$ and an observable $\hat{y}=\int y\mathrm{d}\hat{1}$ in $\mathcal{F}$,
giving $\Pi _{\Delta }$ as the conditional expectation 
\begin{equation*}
\Pi _{\Delta }[C]=\omega _{0}[AE_{\Delta }]\ ,\quad \langle \eta ,\omega
_{0}[X]\eta \rangle =\langle \eta \otimes \varphi _{0},X\eta \otimes \varphi
_{0}\rangle 
\end{equation*}%
of $AE_{\Delta }=U^{\dagger }(C\otimes \hat{1}_{\Delta })U$. In principle,
such a quadruple $(\mathcal{F},\varphi _{0},\hat{y},U)$ was constructed in
[36] and [37] for the normal completely positive $\Pi _{\Delta }$, giving a
justification of the general reduction postulate as described above for the
case of the projective $\Pi _{\Delta }$. For the continuous observation this
problem was solved~[39] on the infinitesimal level in terms of the quantum
stochastic unitary dilation of a differential evolution equation for
characteristic operations 
\begin{equation*}
\tilde{\Psi}(t,q)=\int e^{\mathrm{i}qy}\Psi (t,y)\mathrm{d}\nu \ ,\quad \Psi
(t,y)=\lim_{\Delta \downarrow y}{\frac{1}{\nu _{\Delta }}}\Pi _{\Delta
}^{t}\ , 
\end{equation*}%
where $\mathrm{d}\pi $ is a standard probability measure of $\mathrm{d}%
y\subset \Lambda $. This corresponds to the stationary Markovian evolution
of the convolutional instrumental semigroups $\{\Pi _{\Delta }^{t}|t\geq 0\}$
giving the reduced description of the continuous measurement, with the data $%
y(t)$ having the values in an additive group.

Unfortunately the characteristic operational description of the quantum
measurement is not relevant to the sample-paths representation. It is not
suitable for the conditioning of the quantum evolution under the given data
of the observations and hence does not allow one to obtain explicitly the
corresponding dynamical reduction. Moreover, the continuous measurements
have the data $y$ not necessary in a group, and in the nonstationary cases
they cannot be described by the convolution instrumental semigroups and the
corresponding evolution equations.

Recently a new differential description of continual nondemolition
measurements was developed within the noncommutative stochastic calculus
method [13,14,31]. A general stochastic filtering equation was derived for
the infinitesimal sample-paths representation of the quantum conditional
expectations, giving the continuous generalized reduction of the \emph{a
posteriori\/} states [25,26,29].

Simultaneously, some particular cases of the filtering equation for the
stochastic state-vector $\varphi (t,\omega )=\chi _{y^{t}}(\omega )$,
corresponding to the functional spectrum $\Lambda ^{t}$ of the diffusion
trajectories $y^{t}(\omega )=\{y(s,\omega )|s\leq t\}$, were discovered
within the phenomenological theories of the dynamical reduction and
spontaneous localization [16--18]. As was shown in [21,27] and [29], the
nonlinearity of such equations is related only to the normalization $\Vert
\varphi (t,\omega )\Vert =1$ and after the proper renormalization $\chi
_{t}(\omega )=\sqrt{g_{t}(\omega )}\varphi (t,\omega )$, where $g_{t}(\omega
)$ is the probability density of the process 
\begin{equation*}
y(s,\omega )={\frac{1}{s}}\int_{0}^{s}\langle \varphi (t,\omega ),R\varphi
(t,\omega )\rangle \mathrm{d}t+s^{-1}w_{s}\ ,\quad s\in \lbrack 0,t)
\end{equation*}%
generated by the standard Wiener process $\omega =\{w_{t}\}$ with respect to
the Wiener probability measure $\mathrm{d}\pi $ on the continuous
trajectories $\omega \in \Omega $, they become the linear ones 
\begin{equation}
\mathrm{d}\chi _{t}+\left( {\frac{\mathrm{i}}{\hbar }}H+{\frac{1}{2}}%
L^{\dagger }L\right) \chi _{t}\mathrm{d}t=L\chi _{t}\mathrm{d}w\ .
\label{eq:3.1}
\end{equation}%
Here $H$ is the Hamiltonian of the object, $L$ is an arbitrary operator in $%
\mathcal{H}$ defining the variable $R=\sqrt{\hbar }(L+L^{\dagger })$, under
the continuous measurement, and $\mathrm{d}w=w_{t+\mathrm{d}t}-w_{t}$ is the
forward increment, such that the stochastic equation (\ref{eq:3.1}) has to
be solved in the It\^{o} sense. This solution can be explicitly written as 
\begin{equation}
\chi _{t}(\omega )=T_{t}(\omega )\xi ,\quad T_{t}(\omega )=\exp
\{w_{t}L-tL^{2}\}  \label{eq:3.2}
\end{equation}%
in the case $L=\sqrt{\pi /2h}\,R$, $(h=2\pi \hbar )$, $H=0$, corresponding
to the unsharp measurement of the self-adjoint operator $R$ during the time
interval $[0,t)$ with the trivial free Hamiltonian evolution of the object.
In the case $H\not=0$ this can be used for the approximate solution of (\ref%
{eq:3.1}) with $L^{\dagger }=L$, $\chi (0)=\eta $ as $\chi _{t}(\omega
)\simeq T_{t}(\omega )\xi (t)$, where $\xi (t)=V(t)\eta $ is the unitary
evolution $V(t)=\exp \left\{ -{\frac{\mathrm{i}}{\hbar }}Ht\right\} $
without the measurement.

The stochastic transformation (\ref{eq:3.2}) defines the operational density 
\begin{equation*}
\Theta _{t}[C](\omega )=T_{t}^{\dagger }(\omega )CT_{t}(\omega ) 
\end{equation*}%
of an instrument as in (\ref{eq:2.9}) with respect to the standard Wiener
probability measure $\mathrm{d}\pi $ on $\omega ^{t}=\{w_{s}\}_{s\leq t}\in
\Omega ^{t}$ having the Gaussian marginal distribution of $q_{t}=\sqrt{\hbar 
}w_{t}$ 
\begin{equation*}
\mathrm{d}\nu :=\int_{q_{t}\in \mathrm{d}q}\mathrm{d}\pi =(ht)^{-1/2}\exp
[-\pi q^{2}/ht\}\mathrm{d}q\,. 
\end{equation*}%
Hence $\Psi (t,q)\mathrm{d}\nu :=\int\limits_{q_{t}\in \mathrm{d}q}\Theta
_{t}(\omega )\mathrm{d}\pi =\Phi (t,y)\mathrm{d}y$, where $y={\frac{1}{t}}q$%
, 
\begin{equation}
\Phi \lbrack C](t,y)=\sqrt{\frac{t}{h}}\exp \left\{ -{\frac{\pi t}{2h}}%
(y-R)^{2}\right\} C\exp \left\{ -{\frac{\pi t}{2h}}(y-R)^{2}\right\} \,,
\label{eq:3.3}
\end{equation}%
because $\Theta _{t}(\omega )$ depends only on $w_{t}$: $\Theta _{t}(\omega
)=\Psi (t,\sqrt{\hbar }w_{t})$, and 
\begin{equation*}
\Psi \lbrack C](t,q)=G(t,q)^{\dagger }CG(t,q)\,,\quad G(t,q)=\exp \left\{ -{%
\frac{\pi }{h}}\left( qR-{\frac{t}{2}}R^{2}\right) \right\} \,. 
\end{equation*}%
The operator $E(t,y)=\Phi \lbrack I](t,y)=f_{R}(t,y)$, 
\begin{equation*}
f_{R}(t,y)=\sqrt{\frac{t}{h}}\exp \left\{ -\pi {\frac{t}{h}}%
(y-R)^{2}\right\} =F(t,y)^{\dagger }F(t,y) 
\end{equation*}%
defines the probability density of the unsharp measurement of $R$ with
respect to the ordinary Lebesgue measure $\mathrm{d}y$ as the convolution 
\begin{equation*}
g_{\xi }(t,y)=\int \sqrt{\frac{t}{h}}\exp \left\{ -\pi {\frac{t}{h}}%
(y-x)^{2}\right\} h_{\xi }(x)\mathrm{d}\lambda =(f_{0}\ast h_{\xi })(y)\,, 
\end{equation*}%
where $h_{\xi }(x)=|\xi (x)|^{2}$, $\xi (x)=\langle x|\xi $, $\mathrm{d}%
\lambda =\sum \delta (x-x_{i})\mathrm{d}x$ in the case of discrete spectrum $%
\{x_{i}\}$ of $R$, and $\mathrm{d}\lambda =\mathrm{d}x$ in the case of
purely continuous spectrum of $R$.

This means that the continuous unsharp measurement of $R$ can be described
by the observation model $y_x(t)=x+(1/t)q_t$ of signal $x$ plus Gaussian
error $e(t)=(1/t)q_t$ with independent increments as 
\begin{equation}
y_R(t)=R+e(t)I\,,\quad e(t)={\frac{\sqrt\hbar}{t}}w_t\,.  \label{eq:3.4}
\end{equation}
The noise $e(t)$ with the mean value $\langle e(t)\rangle=0$ gives a
decreasing unsharpness $\langle e(t)^2\rangle=\hbar/t$ of the measurement
from infinity to zero that is inversely proportional to the duration of the
observation interval $t>0$.

In general, such a model can be realized [21]--[25] \nocite{bib:21,22,23}%
\nocite{bib:24,25} as the nondemolition observation within the quantum
stochastic theory of unitary evolution of the compound system on the product 
${\mathcal{G}=\mathcal{H}}\otimes \mathcal{F}$ with the Fock space $\mathcal{%
F}$ over the one-particle space $L^{2}(\mathbb{R}_{+})$ for a
one-dimensional bosonic field, modeling the measurement apparatus of the
continuous observation.

Let us illustrate this general construction for our particular case $H=0$, $%
L=L^{\dagger }$. The unitary interaction $S(t)$ in $\mathcal{G}$, defining
the transformations (\ref{eq:3.2}) as (\ref{eq:2.2}) with respect to the
vacuum state-vector $\varphi _{0}\in \mathcal{F}$, is generated by the field
momenta operators 
\begin{equation}
\hat{p}_{s}={\frac{\mathrm{i}}{2}}\sqrt{\hbar }(\hat{a}_{s}^{\dagger }-\hat{a%
}_{s})\,,\quad s\in \mathbb{R}_{+}  \label{eq:3.5}
\end{equation}%
as $S(t)=\exp \left\{ -{\frac{\mathrm{i}}{\hbar }}R\otimes \hat{p}%
_{t}\right\} $.

Here $\hat{a}_{s}$ and $\hat{a}_{s}^{\dagger }$ are the canonical
annihilation and creation operators in $\mathcal{F}$, localized on the
intervals $[0,s]$ according to the commutation relations 
\begin{equation*}
\lbrack \hat{a}_{r},\hat{a}_{s}]=0,\quad \lbrack \hat{a}_{r},\hat{a}%
_{s}^{\dagger }]=s\hat{1}\ ,\quad \forall r\geq s\,, 
\end{equation*}%
The pointer of the apparatus for the measurement of $R$ is defined by the
field coordinate observables 
\begin{equation}
\hat{q}_{s}=\sqrt{\hbar }(\hat{a}_{s}+\hat{a}_{s}^{\dagger })\,,\quad s\in 
\mathbb{R}_{+}  \label{eq:3.6}
\end{equation}%
which are compatible with $[\hat{q}_{r},\hat{q}_{s}]=0$ as well as with $[%
\hat{p}_{r},\hat{p}_{s}]=0$, but incompatible with (\ref{eq:3.5}): 
\begin{equation*}
\lbrack \hat{p}_{r},\hat{q}_{s}]=s{\frac{\hbar }{\mathrm{i}}}\hat{1}\ ,\quad
\forall r\geq s\,. 
\end{equation*}%
The operators $S^{\dagger }(t)$ satisfy the condition (\ref{eq:1.2}): $%
\langle x|S(t)=\hat{s}_{x}(t)\langle x|$, where the unitary operators $\hat{s%
}_{x}^{\dagger }(t):\mathcal{F}\rightarrow \mathcal{F}$ can be described by
the shifts 
\begin{equation}
\hat{s}_{x}^{\dagger }(t):|q,t\rangle \mapsto |xt+q,t\rangle \,,\quad
\forall x,q,t
\end{equation}%
similarly to (\ref{eq:2.7}). Here $|q,t\rangle $ is the (generalized)
marginal eigenvector of the self-adjoint operator 
\begin{equation*}
\hat{e}(t)=t^{-1}\hat{q}_{t}\ ,\quad \hat{q}_{t}|q,t\rangle =q|q,t\rangle
\,, 
\end{equation*}%
uniquely defined up to the phase by an eigenvalue $q\in \mathbb{R}$ as the
Dirac $\delta $-function $\delta _{q}$ in the $\hat{q}_{t}$-representation $%
L^{2}(\mathbb{R})$ of the Hilbert subspace $\mathcal{A}(t)\varphi
_{0}\subset \mathcal{F}$, where $\mathcal{A}(t)$ is the Abelian algebra
generated by $\hat{q}_{t}$, and $\varphi _{0}\in \mathcal{F}$ is the
vacuum--vector of the Fock space $\mathcal{F}$. Due to this, 
\begin{equation*}
\hat{y}_{x}(t)=\hat{s}_{x}^{\dagger }(t)\hat{e}(t)\hat{s}_{x}(t)=x\hat{1}+%
\hat{e}(t)\,, 
\end{equation*}%
which gives the quantum stochastic realization of the observation model (\ref%
{eq:3.4}) in terms of the output nondemolition process $\hat{y}_{R}(t)={%
\frac{1}{t}}Y(t)$, 
\begin{equation}
Y(t)=S^{\dagger }(t)(I\otimes \hat{q}_{t})S(t)=tR\otimes \hat{1}+I\otimes 
\hat{q}_{t}  \label{eq:3.8}
\end{equation}%
similarly to (\ref{eq:2.8}) with $\hat{q}_{t}$ represented by the operator $%
\sqrt{\hbar }(\hat{a}_{t}+\hat{a}_{t}^{\dagger })$. Indeed, the classical
noise $q_{t}=\sqrt{\hbar }w_{t}$ is statistically equivalent to the quantum
one $\hat{q}_{t}=\sqrt{\hbar }(\hat{a}_{t}+\hat{a}_{t}^{\dagger })$ with
respect to the vacuum state, as can be seen by a comparison of their
characteristic functionals:%
\begin{eqnarray*}
\langle e^{\mathrm{i}\int f(s)\mathrm{d}q}\rangle &:&=\int \exp \{\mathrm{i}%
\sqrt{\hbar }\int_{0}^{\infty }f(s)\mathrm{d}w\}\mathrm{d}\pi =\exp \left\{ -%
{\frac{\hbar }{2}}\int_{0}^{\infty }f(s)^{2}\mathrm{d}s\right\} \\
&=&\langle \varphi _{0},e^{\mathrm{i}\int f(s)\mathrm{d}\hat{a}^{\dagger
}}e^{-{\frac{\hbar }{2}}\int_{0}^{\infty }f(s)^{2}\mathrm{d}s}e^{\mathrm{i}%
\int f(s)\mathrm{d}\hat{a}}\varphi _{0}\rangle =\langle \varphi _{0},e^{%
\mathrm{i}\int f(s)\mathrm{d}\hat{q}}\varphi _{0}\rangle \,.
\end{eqnarray*}%
Here we used the annihilation property $\hat{a}_{s}\varphi _{0}=0$ and the
Wick ordering formula 
\begin{equation}
\exp \{z^{\prime }\hat{a}_{s}+\hat{a}_{s}^{\dagger }z\}=e^{z\hat{a}%
_{s}^{\dagger }}\exp \left\{ z^{\prime }{\frac{s}{2}}z\right\} e^{z^{\prime }%
\hat{a}_{s}}\ .  \label{eq:3.9}
\end{equation}%
The observable process (\ref{eq:3.8}) satisfies the nondemolition condition (%
\ref{eq:1.1}) (and self-nondemolition) with respect to any quantum process $%
X(t)=\left( S^{\dagger }ZS\right) (t)$ given by the operators $Z(t)$,
commuting with all $Q(s)=I\otimes \hat{q}(s)$, $s\leq t$, because 
\begin{equation*}
Y(s)=S^{\dagger }(t)(I\otimes \hat{q}(s))S(t)\,,\quad \forall s\leq t\ , 
\end{equation*}%
as follows from the commutation relations 
\begin{equation*}
\hat{s}_{x}^{\dagger }(t)\hat{q}_{s}=(sx\hat{1}+\hat{q}_{s})\hat{s}%
_{x}^{\dagger }(t)\ ,\quad \forall s\leq t 
\end{equation*}%
for $\hat{s}_{x}^{\dagger }(t)=\exp \left\{ {\frac{\mathrm{i}}{\hbar }}x\hat{%
p}_{t}\right\} $. Indeed, due to this 
\begin{equation*}
\lbrack X(t),Y(s)]=W(t)[Z(t),Q(s)]W^{\dagger }(t)=0\,, 
\end{equation*}%
if $t>s$ and $[Z(t),Q(s)]=0$, as in the cases $Z(t)=C\otimes \hat{1}$ and $%
Z(t)=Q(t)$, where $Q(t)=I\otimes \hat{q}_{t}$.

Now we can find the transform 
\begin{equation*}
\langle q,t|S\varphi _{0}=G(t,q)\varphi _{0}(t,q)={\frac{1}{\sqrt{t}}}%
T\left( t,{\frac{1}{t}}q\right) \,, 
\end{equation*}%
where $\varphi _{0}(t,q)=\langle q,t|\varphi _{0}$ is the vacuum-vector $%
\varphi _{0}\in \mathcal{F}$ in the marginal $\hat{q}_{t}=q$ representation 
\begin{equation*}
\varphi _{0}(t,q)=(ht)^{-1/4}\exp \{-\pi q^{2}/2ht\}\,,\quad q\in \mathbb{R} 
\end{equation*}%
normalized with respect to the Lebesgue measure $\mathrm{d}q$ on $\mathbb{R}$%
. To this end, let us apply the formula (\ref{eq:3.9}) to $S^{\dagger
}(t)=\exp \left\{ {\frac{\mathrm{i}}{\hbar }}R\otimes \hat{p}_{t}\right\} $: 
\begin{equation*}
\exp \{-L\otimes \hat{a}_{t}+L\otimes \hat{a}_{t}^{\dagger }\}=e^{L\otimes 
\hat{a}_{t}^{\dagger }}\exp \left\{ -{\frac{t}{2}}L^{2}\right\} e^{-L\otimes 
\hat{a}_{t}}\,, 
\end{equation*}%
where $L=R/2\sqrt{\hbar }$. Using the annihilation property $\exp \{\pm
L\otimes \hat{a}_{t}\}\varphi _{0}=\varphi _{0}$, we obtain 
\begin{eqnarray*}
W(t)^{\dagger }\varphi _{0} &=&e^{L\otimes \hat{a}_{t}^{\dagger }}\exp
\left\{ -{\frac{t}{2}}L^{2}\right\} e^{-L\otimes \hat{a}_{t}}\varphi _{0} \\
&=&e^{L\otimes \hat{a}_{t}^{\dagger }}\exp \left\{ -{\frac{t}{2}}%
L^{2}\right\} e^{L\otimes \hat{a}_{t}}\varphi _{0}=e^{L\otimes \hat{w}%
_{t}-tL^{2}}\varphi _{0}\,.
\end{eqnarray*}%
This is equivalent to (\ref{eq:3.2}) because of the Segal isometry of the
vectors $\exp \{x\hat{w}_{t}\}\varphi _{0}\in \mathcal{F}$, where $x\in 
\mathbb{R} $, $\hat{w}_{t}=\hat{a}_{t}+\hat{a}_{t}^{\dagger }$, and the
stochastic functions $\exp \{xw_{t}\}\in L_{\pi }^{2}(\Omega )$ in the
Hilbert space of the Wiener measure $\pi $ on $\Omega $. Hence the transform 
$F\left( t,{\frac{1}{t}}q\right) =\sqrt{t}G(t,q)\varphi _{0}(t,q)$ defining
the density $\Phi (t,y)=F(t,y)^{\dagger }[\cdot ]F(t,y)$ of the instrument (%
\ref{eq:2.9}) with respect to $\mathrm{d}y$ has the same form, as in (\ref%
{eq:3.3}): 
\begin{equation}
F(t,y)=(t/h)^{1/4}\exp \left\{ -{\frac{\pi t}{2h}}(y-R)^{2}\right\} \,.
\label{eq:3.10}
\end{equation}

\section{A Hamiltonian model for continuous reduction}

As we have shown in the previous section the continuous reduction equation (%
\ref{eq:3.1}) for the non-normalized stochastic state-vector $\chi(t,\omega)$
can be obtained from an interaction model of the object of measurement with
a bosonic field. This can be done by conditioning with respect to a
nondemolition continuous observation of field coordinate observables (\ref%
{eq:3.6}) in the vacuum state.

The unitary evolution $\psi (t)=U(t)\psi _{0}$ in the tensor product $%
\mathcal{G}=\mathcal{H}\otimes \mathcal{F}$ with the Fock space $\mathcal{F}$
corresponding to (\ref{eq:3.1}) can be written as the generalized Schr\"{o}%
dinger equation 
\begin{equation}
\mathrm{d}\psi (t)+K_{0}\psi (t)\mathrm{d}t=(L\otimes \mathrm{d}\hat{a}%
_{t}^{\dagger }-L^{\dagger }\otimes \mathrm{d}\hat{a}_{t})\psi (t)
\label{eq:4.1}
\end{equation}%
in terms of the annihilation and creation canonical field operators $\hat{a}%
_{s}$, $\hat{a}_{s}^{\dagger }$. This is a singular differential equation
which has to be treated as a quantum stochastic one [29] in terms of the
forward increments $\mathrm{d}\psi (t)=\psi (t+\mathrm{d}t)-\psi (t)$ with $%
K_{0}=K\otimes \hat{1}$, $K=(\mathrm{i}/\hbar )H+{\frac{1}{2}}L^{\dagger }L$%
. In the particular case $L=R/2\sqrt{\hbar }=L^{\dagger }$ of interest, eq. (%
\ref{eq:4.1}) can be written simply as a classical stochastic one, $\mathrm{d%
}\psi +K\psi \mathrm{d}t=(\mathrm{i}/\hbar )R\mathrm{d}p$, in It\^{o} sense
with respect to a Wiener process $p_{t}$ of the same intensity $(\mathrm{d}%
p_{t})^{2}=\hbar \mathrm{d}t/4$ as the field momenta operators (\ref{eq:3.5}%
) with respect to the vacuum state. But the standard Wiener process $%
v_{t}=2p_{t}/\sqrt{\hbar }$ cannot be identified with the Wiener process $%
w_{t}$ in the reduction equation (\ref{eq:3.1}) because of the nondemolition
principle. Moreover, there is no way to get the nondemolition property (\ref%
{eq:1.1}) for 
\begin{equation*}
X(t)=U(t)^{\dagger }X_{0}U(t)\ ,\quad Y(s)=U(s)^{\dagger }Y_{0}(s)U(s)
\end{equation*}%
with the independent or if only commuting $v_{t}$ and $w_{t}$, as one can
see in the simplest case $H=0$, $X_{0}={\frac{\hbar }{\mathrm{i}}}{\frac{%
\mathrm{d}}{\mathrm{d}x}}\otimes \hat{1}$, $R=x$, $Y_{0}(s)=I\otimes \hat{q}%
_{s}$.

Indeed, the error process $q_{t}=\sqrt{\hbar }w_{t}$ is appearing in (\ref%
{eq:3.4}) as a classical representation of the field coordinate observables (%
\ref{eq:3.6}) which do not commute with (\ref{eq:3.5}). In this case, eq. (%
\ref{eq:4.1}) gives the unitary operator $U(t)=\exp \left\{ -{\frac{\mathrm{i%
}}{\hbar }}x\otimes \hat{p}_{t}\right\} $ and the Heisenberg operators 
\begin{equation*}
X(t)={\frac{\hbar }{\mathrm{i}}}{\frac{\mathrm{d}}{\mathrm{d}x}}\otimes \hat{%
1}-I\otimes \hat{p}_{t}\ ,\quad Y(s)=sx\otimes \hat{1}+I\otimes \hat{q}_{s} 
\end{equation*}%
commute for all $t\geq s$ only because 
\begin{equation*}
\left[ {\frac{\hbar }{\mathrm{i}}}{\frac{\mathrm{d}}{\mathrm{d}x}},sx\right]
\otimes \hat{1}=s{\frac{\hbar }{\mathrm{i}}}I\otimes \hat{1}=[\hat{p}_{t},%
\hat{q}_{s}]\ ,\quad \forall t\geq s\,. 
\end{equation*}%
Hence, there is no way to obtain (\ref{eq:1.1}) for the classical stochastic
processes $p_{t}$, $q_{s}$ by replacing simultaneously $\hat{p}_{t}$ and $%
\hat{q}_{s}$ for commuting $\sqrt{\hbar }v_{t}/2$ and $\sqrt{\hbar }w_{t}$
even though $p_{t}$ is statistically identical to $\hat{p}_{t}$ and
separately $q_{s}$ to $\hat{q}_{s}$.

Let us show now how one can get a completely different type of the reduction
equation than postulated in [16]--[20] simply by fixing an another
nondemolition process for the same interaction, corresponding to the Schr%
\"{o}dinger stochastic equation (\ref{eq:4.1}) with $L=L^{\dagger }$ and $%
H=0 $.

We fix the discrete pointer of the measurement apparatus, which is described
by the observable $\hat{n}_{s}={\frac{1}{s}}\hat{a}_{s}^{\dagger }\hat{a}_{s}
$, by counting the quanta of the Bosonic field in the mode $1_{s}(r)=1$ if $%
r\in \lbrack 0,s)$ and $1_{s}(r)=0$ if $r\notin \lbrack 0,s)$. The operators 
$\hat{n}_{t}$ have the integer eigenvalues $0,1,2,\dots $ corresponding to
the eigen-vectors 
\begin{equation*}
|n,t\rangle =e^{t/2}(\hat{a}_{t}^{\dagger }/t)^{n}\varphi _{0}\ ,\quad \hat{a%
}_{t}\varphi _{0}=0
\end{equation*}%
which we have normalized with respect to the standard Poissonian
distribution 
\begin{equation}
\nu _{n}=e^{-t}t^{n}/n!\ ,\quad n=0,1,\dots \   \label{eq:4.2}
\end{equation}%
as $\langle n,t|n,t\rangle =1/\nu _{n}$. Let us find the matrix elements 
\begin{equation*}
\langle n,t|S(t)\varphi _{0}=G(t,n)
\end{equation*}%
for the unitary evolution operators 
\begin{equation}
S(t)=\exp \{-L\otimes \hat{a}_{t}+L\otimes \hat{a}_{t}^{\dagger }\}\,,
\label{eq;4.3}
\end{equation}%
by resolving eq. (\ref{eq:4.1}) in the considered case. This can be done
again by representing $S(t)$ in the form (\ref{eq:3.9}) for $z^{\prime }=L$, 
$z=-L$ and the commutation rule 
\begin{equation*}
(I\otimes \hat{a}_{t})e^{L\otimes \hat{a}_{t}^{\dagger }}=e^{L\otimes \hat{a}%
_{t}^{\dagger }}(tL\otimes \hat{1}+I\otimes \hat{a}_{t})\ .
\end{equation*}%
Due to the annihilation property, this gives 
\begin{equation}
\varphi _{0}^{\dagger }(\hat{a}_{t}/t)^{n}e^{L\otimes \hat{a}_{t}^{\dagger
}}\exp \left\{ {\frac{t}{2}}(1-L^{2})\right\} e^{-L\otimes \hat{a}%
_{t}}\varphi _{0}=L^{n}\exp \left\{ {\frac{t}{2}}(1-L^{2})\right\} =G(t,n)\ .
\label{eq:4.4}
\end{equation}%
The obtained reduction transformations are not unitary and not projective
for any $n=0,1,2,\dots $, but they define the nonorthogonal identity
resolution 
\begin{equation*}
\sum_{n=0}^{\infty }G(t,n)^{\dagger }G(t,n)e^{-t}t^{n}/n!=I
\end{equation*}%
corresponding to the operational density 
\begin{equation}
\Psi \lbrack C](t,n)=e^{t}L^{n}e^{-L^{2}/2}CL^{n}e^{-L^{2}/2}  \label{eq:4.5}
\end{equation}%
with respect to the measure (\ref{eq:4.2}). Now we can easily obtain the
stochastic reduction equation for $\chi (t,\omega )=T(t,\omega )\eta $ if we
replace the eigenvalue $n$ of $\hat{n}_{t}$ by the standard Poissonian
process $n_{t}(\omega )$ with the marginal distributions (\ref{eq:4.2}).
Such a process $n_{t}$ describes the trajectories $t\mapsto n_{t}(\omega )$
that spontaneously increase by $\mathrm{d}n_{t}(\omega )=1$ at random time
instants $\omega =\{t_{1}<t_{2}<\dots \}$ as the spectral functions $%
\{n_{t}(\omega )\}$ for the commutative family $\{\hat{n}_{t}\}$. The
corresponding equation for the stochastic state-vector $\chi (t,\omega
)=\chi (t,n_{t}(\omega ))$ can be written in the It\^{o} sense as 
\begin{equation}
\mathrm{d}\chi (t)+{\frac{1}{2}}(L^{2}-I)\chi (t)\mathrm{d}t=(L-1)\chi (t)%
\mathrm{d}n_{t}\,.  \label{eq:4.6}
\end{equation}%
Obviously Eq. (\ref{eq:4.6}) has the unique solution $\chi (t)=F(t)\eta $
written for a given $\eta \in \mathcal{H}$ as 
\begin{equation}
\chi (t)=L^{n_{t}}\exp \left\{ {\frac{t}{2}}(1-L^{2})\right\} \eta
=G(t,n_{t})\eta   \label{eq:4.7}
\end{equation}%
because of $\mathrm{d}\chi (t)=(L-1)\chi (t)$ when $\mathrm{d}n_{t}=1$,
otherwise $\mathrm{d}\chi (t)={\frac{1}{2}}(1-L^{2})\chi (t)\mathrm{d}t$ in
terms of the forward differential $\mathrm{d}\chi (t)=\chi (t+\mathrm{d}%
t)-\chi (t)$.

Such an equation was derived in [26]--[30] also for the general quantum
stochastic equation (\ref{eq:4.1}) on the basis of quantum stochastic
calculus and filtering theory [31]. Moreover, it was proved that any other
stochastic reduction equation can be obtained as a mixture of Eq. (\ref%
{eq:3.1}) and (\ref{eq:4.4}) which are of fundamentally different types.

Finally let us write down a Hamiltonian interaction model corresponding to
the quantum stochastic Schr\"{o}dinger equation (\ref{eq:4.1}). Using the
notion of chronologically ordered exponential 
\begin{equation}
U(t)=\exp ^{\leftarrow }\left\{ -{\frac{\mathrm{i}}{\hbar }}\int_{0}^{t}H(r)%
\mathrm{d}r\right\}  \label{eq:4.8}
\end{equation}%
one can extend its solutions $\psi (t)=\exp \left\{ -{\frac{\mathrm{i}}{%
\hbar }}R\otimes \hat{p}_{t}\right\} \psi _{0}$ also to the general case, $%
H\not=0$, $L^{\dagger }\not=L$ in terms of the generalized Hamiltonian 
\begin{equation*}
H(t)=H_{0}+{\frac{\hbar }{\mathrm{i}}}(L^{\dagger }\otimes \hat{a}%
(t)-L\otimes \hat{a}(t)^{\dagger })\,, 
\end{equation*}%
where $\hat{a}(t)=\mathrm{d}\hat{a}_{t}/\mathrm{d}t$, $\hat{a}^{\dagger }(t)=%
\mathrm{d}\hat{a}_{t}^{\dagger }/\mathrm{d}t$, $H_{0}=H\otimes \hat{1}$. The
time-dependent Hamiltonian $H(t)$ can be treated as the object interaction
Hamiltonian 
\begin{equation*}
H(t)=H_{0}+{\frac{\hbar }{\mathrm{i}}}e^{{\frac{\mathrm{i}}{\hbar }}%
H_{1}t}(L^{\dagger }\otimes \hat{a}(0)-L\otimes \hat{a}(0)^{\dagger })e^{-{%
\frac{\mathrm{i}}{\hbar }}H_{1}t} 
\end{equation*}%
for a special free evolution Hamiltonian $H_{1}=I\otimes \hat{h}$ of the
quantum bosonic field $\hat{a}(r)$, $r\in \mathbb{R}$ described by the
canonical commutation relations 
\begin{equation*}
\lbrack \hat{a}(r),\hat{a}(s)]=0,\quad \lbrack \hat{a}(r),\hat{a}%
(s)^{\dagger }]=\delta (r-s)\hat{1}\,,\quad \forall \,r,s\in \mathbb{R}\,. 
\end{equation*}%
This free evolution in the Fock space $\mathcal{F}$ over one particle space $%
L^{2}(\mathbb{R})$ is simply given by the shifts 
\begin{equation*}
e^{{\frac{\mathrm{i}}{\hbar }}\hat{h}t}\hat{a}(r)e^{-{\frac{\mathrm{i}}{%
\hbar }}\hat{h}t}=\hat{a}(r+t)\,,\quad \forall \,r,t\in \mathbb{R}\,, 
\end{equation*}%
corresponding to the second quantization $\hat{h}=\hat{a}^{\dagger }\hat{%
\varepsilon}\hat{a}$ of the one-particle Hamiltonian $\hat{\varepsilon}={%
\frac{\hbar }{\mathrm{i}}}{\frac{\partial }{\partial r}}$ in $L^{2}(\mathbb{R%
})$. Hence, the total Hamiltonian of the system \textquotedblleft object
plus measurement apparatus\textquotedblright\ can be written as 
\begin{equation}
H_{s}=H\otimes \hat{1}+{\frac{\hbar }{\mathrm{i}}}(L^{\dagger }\otimes \hat{a%
}(0)-L\otimes \hat{a}(0)^{\dagger }+I\otimes \hat{a}^{\dagger }\hat{a}%
^{\prime })\,,  \label{eq:4.9}
\end{equation}%
where $a^{\dagger }a^{\prime }=\int\limits_{-\infty }^{\infty }\hat{a}%
(r)^{\dagger }\hat{a}(r)^{\prime }$\textrm{$d$}$r$, $\hat{a}(r)^{\prime }=%
\mathrm{d}\hat{a}(r)/\mathrm{d}r$. Of course, the free field Hamiltonian $%
\hat{h}=\hbar \hat{a}^{\dagger }\hat{a}^{\prime }/\mathrm{i}$ is rather
unusual as with respect to the single-particle energy $\varepsilon (p)=p$ in
the momentum representation giving the unbounded (from below) spectrum of $%
\hat{\varepsilon}$.

But one can consider such an energy as an approximation 
\begin{equation}
\varepsilon (p)=\lim_{p_{0}\rightarrow \infty }c\left( \sqrt{%
(p+p_{0})^{2}+(m_{0}c)^{2}}-\sqrt{p_{0}^{2}+(m_{0}c)^{2}}\right) =v_{0}p
\label{eq:4.10}
\end{equation}%
in the velocity units $v_{0}=c/\sqrt{1+(m_{0}c/p_{0})^{2}}=1$ for the shift $%
\varepsilon _{0}(p)-\varepsilon _{0}(0)$ of the standard relativistic energy 
$\varepsilon _{0}(p)=c\sqrt{(p+p_{0})^{2}+(m_{0}c)^{2}}$ as the function of
small deviations $|p|\ll p_{0}$ from the initially fixed momentum $p_{0}>0$.
This corresponds to the treatment of the measurement apparatus as a beam of
bosons with mean momentum $p_{0}\rightarrow \infty $ given in an initial
coherent state by a plane wave 
\begin{equation*}
f_{0}(r)=c\exp \{\mathrm{i}p_{0}r/\hbar \}\,. 
\end{equation*}%
This input beam of bosons illuminate the position $R=\sqrt{\hbar }%
(L+L^{\dagger })$ of the object of measurement via the observation of the
commuting position operators $Y(t)$, $t\in \mathbb{R}$ of the output field
given by the generalized Heisenberg operator-process, 
\begin{eqnarray*}
\dot{Y}(t) &=&e^{{\frac{\mathrm{i}}{\hbar }}H_{s}t}(I\otimes \hat{q}(0))e^{-{%
\frac{\mathrm{i}}{\hbar }}H_{s}t} \\
&=&U(t)^{\dagger }(I\otimes \hat{q}(t))U(t)=R(t)+I\otimes \hat{q}(t)
\end{eqnarray*}%
This is the simplest quantum Hamiltonian model for the continuous
nondemolition measurement of the physical quantity $R$ of a quantum object.

Thus the unitary evolution group $U_{s}(t)=e^{{\frac{-\mathrm{i}}{\hbar }}%
H_{s}t}$ of the compound system is defined on the product $\mathcal{H}%
\otimes \mathcal{F}$ with the two--sided Fock space $\mathcal{F}=\Gamma
(L^{2}(\mathbb{R}))$ by $U_{s}(t)=V_{1}(t)U(t)$, where $V_{1}=I\otimes \hat{v%
}$ is the free evolution group $\hat{v}(t)=e^{{\frac{-\mathrm{i}}{\hbar }}%
\hat{h}t}$ of the field, corresponding to the shifts 
\begin{equation*}
f\in L^{2}(\mathbb{R})\mapsto f^{t}(s)=f(s-t) 
\end{equation*}%
of the one-particle space $L^{2}(\mathbb{R})$. To obtain such an evolution
from a realistic Hamiltonian of a system of atoms interacting with an
electromagnetic field one has to use a Markovian approximation,
corresponding to the weak-coupling or low density limits [39].

Thus, the problem of unitary dilation of the continuous reduction and
spontaneous collapse was solved in [25] even for infinite-dimensional Wiener
noise in a stochastic equation of type (\ref{eq:3.2}).

\section*{Conclusion}

Analysis [1] of the quantum measurement notion shows that it is a complex
process, consisting of the stage of preparation [15] and the stage of
registration, i.e., fixing of the pointer and its output state and the
objectification [40].

The dynamical process of the interaction is properly treated within the
quantum theory of singular coupling to get the nontrivial models of
continuous nondemolition observation while the statistical process of the
objectification is properly treated within the quantum theory of stochastic
filtering to get the nonlinear models of continuous spontaneous localization
[21--31].

The nondemolition principle plays the role of superselection for the
observable processes provided the quantum dynamics is given and restricts
the dynamics provided the observation is given. It is a necessary and
sufficient condition for the statistical interpretation of quantum
causality, giving rise to the quantum noise environment but not to the
classical noise environment of the phenomenological continuous reduction and
spontaneous localization theories [16--20].

The axiomatic quantum measurement theory based on the nondemolition
principle abandons the projection postulate as the redundancy given by a
unitary interaction with a meter in the initial eigen-state. It treats the
reduction of the wave packet not as a real dynamical process but as the
statistical evaluation of the \emph{a posteriori\/} states for the
prediction of the probabilities of the future measurements conditioned by
the past observation.

There is no need to postulate a nonstandard, nonunitary, and nonlinear
evolution for the continuous state-vector reduction in the phenomenological
quantum theories of spontaneous localization, and there is no universal
reduction modification of the fundamental Schr\"odinger equation. The
nonunitary stochastic evolution giving the continuous reduction and the
spontaneous localization of the state-vector can be and has been rigorously
derived within the quantum stochastic theory of unitary evolution of the
corresponding compound system, the object of the measurement and an input
Bose field in the vacuum state.

The statistical treatment of the quantum measurement as nondemolition
observation is possible only in the framework of open systems theory in the
spirit of the modern astrophysical theory of the spreading universe. The
open systems theory assumes the possibility of producing for each quantum
object an arbitrary time series of its copies and enlarges these objects
into an environment, a quantum field, innovating the measurement apparatus
by means of a singular interaction for a continuous observation.

It is nonsense to consider seriously a complete observation in the closed
universe; there is no universal quantum observation, no universal reduction
and spontaneous localization for the wave function of the world. Nobody can
prepare an\textit{\ a priori\/ }state compatible with a complete world
observation and reduce the \textit{a posteriori}\emph{\/} state, except God.
But acceptance of God as an external subject of the physical world is at
variance with the closeness assumption of the universe. Thus, the world
state-vector has no statistical interpretation, and the humanitarian
validity of these interpretations would, in any case, be zero. The
probabilistic interpretation of the state-vector is relevant to only the
induced states of the quantum open objects being prepared by
experimentalists in an appropriate compound system for the nondemolition
observation to produce the reduced states after the registration.

\section*{Acknowledgment}

This work was supported by Deutsche Forschung Gemeinschaft at Philipps
Universit\"at, Marburg. I am deeply grateful to Professors L. Accardi, O.
Melsheimer, and H. Neumann for stimulating discussions and encouragement.

\section*{References}

\begin{description}
\item {[1]} \textsc{G. Ludwig}, \textit{Math. Phys.}, 4:331 (1967); 9, 1
(1968).

\item {[2]} \textsc{E.B. Davies, J. Lewis}, \textit{Commun. Math. Phys.},
17:239--260, (1970).

\item {[3]} \textsc{L.E. Ballentine}, \textit{Rev. Mod. Phys.}, 42:358--381,
(1970).

\item {[4]} \textsc{A. Shimony}, \textit{Phys. Rev. D}, 9:2321--2323, (1974).

\item {[5]} \textsc{V.P. Belavkin}, Optimal linear random filtration of
quantum Boson signals. Problems of Control and Inform. Theory, 3:47--62,
(1974).

\item {[6]} \textsc{V.P. Belavkin}, Optimal quantum filtration of Markovian
singals. Problems of Control and Inform. Theory", 7(5):345--360, (1978).

\item {[7]} \textsc{V.P. Belavkin}, Optimal filtering of Markov signals with
quantum noise, \textit{Radio Eng. Electron. Physics}, 25:1445--1453, (1980).

\item {[8]} \textsc{A. Barchielli, L. Lanz, G.M. Prosperi}, \textit{Nuovo
Cimento}, 72B:79, (1982).

\item {[9]} \textsc{V.P. Belavkin}, Theory of Control of Observable Quantum
Systems, \textit{Automatica and Remote Control}, 44(2):178--188, (1983).

\item {[10]} \textsc{A. Peres}, \textit{Am. J. Phys.}, 52:644, (1984).

\item {[11]} \textsc{R.L. Stratonovich}, Conditional Markov processes and
their applications to optimal control, \textit{MGU}, Moscow 1966.

\item {[12]} \textsc{R.E. Kalman, R.S. Bucy}, New results in linear
filtering theory and prediction problems, \textit{J. Basic Engineering,
Trans. ASME}, 83:95--108, (1961).

\item {[13]} \textsc{V.P. Belavkin}, Nondemolition measurement and control
in quantum dynamical systems. In: Proc. of CISM seminar on "\textit{Inform.
Compl. and Control in Quantum Physics\textquotedblright }, A. Blaquiere,
ed., Udine 1985, 311--239, Springer--Verlag, Wien, 1987.

\item {[14]} \textsc{V.P. Belavkin}, Nondemolition measurements, nonlinear
filtering and dynamical programming of quantum stochastic processes. In:
Proc. of Bellmann Continuum Workshop \textquotedblleft \textit{Modelling and
Control of Systems}\textquotedblright , A. Blaquiere, ed., Sophia--Antipolis
1988, 245--265, \textit{Lect. Not. Cont. Inf. Sci.}, 121, Springer--Verlag,
Berlin, 1988.

\item {[15]} \textsc{L.E. Ballentine}, \textit{Int. J. Theor. Phys.},
27:211--218, (1987).

\item {[16]} \textsc{P. Pearle}, \textit{Phys. Rev.}, D29:235, (1984).

\item {[17]} \textsc{N. Gisen}, \textit{J. Phys. A.: Math. Gen.},
19:205--210, (1986).

\item {[18]} \textsc{L. Diosi}, \textit{Phys. Rev.}, A40:1165--1174, (1988).

\item {[19]} \textsc{G.C. Ghirardi, A. Rimini, T. Weber}, \textit{Phys. Rev.}%
. D34(2):470--491, (1986).

\item {[20]} \textsc{G.C. Ghirardi, P. Pearle, A. Rimini}, \textit{Phys. Rev.%
}, A42:478--89, (1990).

\item {[21]} \textsc{V.P. Belavkin}, A new wave equation for a continuous
non--demolition measurement", \textit{Phys. Lett.}, A140:355--358, (1989).

\item {[22]} \textsc{V.P. Belavkin, P. Staszewski}, A quantum particle
undergoing continuous observation", \textit{Phys. Lett.}, A140:359--362,
(1989).

\item {[23]} \textsc{V.P. Belavkin}, A posterior Schr\"{o}dinger equation
for continuous non--demolition measurement, \textit{J. Math. Phys},
31(12):2930--2934, (1990).

\item {[24]} \textsc{V.P. Belavkin, P. Staszewski}, Nondemolition
observation of a free quantum particle, \textit{Phys. Rev. A.\/},
45(3):1347--1356, (1992).

\item {[25]} \textsc{V.P. Belavkin}, Quantum continual measurements and a
posteriori collapse on CCR, \textit{Commun. Math. Phys.}, 146, 611--635,
(1992).

\item {[26]} \textsc{V.P. Belavkin}, A continuous counting observation and
posterior quantum dynamics, \textit{J. Phys. A, Math. Gen.}, 22: L
1109--1114, (1989).

\item {[27]} \textsc{V.P. Belavkin}, A stochastic posterior Schr\"{o}dinger
equation for counting non--demolition measurement, \textit{Letters in Math.
Phys.}, 20"85--89, (1990).

\item {[28]} \textsc{V.P. Belavkin, P. Staszewski}, \textit{Rep. Math. Phys.}%
, 29:213--225, (1991).

\item {[29]} \textsc{V.P. Belavkin}, Stochastic posterior equations for
quantum nonlinear filtering. Probab., \textit{Theory and Math. Stat.}, ed.
B. Grigelionis, 1:91--109, VSP/Mokslas 1990.

\item {[30]} \textsc{A. Barchielli, V.P. Belavkin}, Measurements continuous
in time and a posteriori states in quantum mechanics, \textit{J. Phys. A,
Math. Gen.}, 24:1495--1514, (1991).

\item {[31]} \textsc{V.P. Belavkin}, Quantum stochastic calculus and quantum
nonlinear filtering, \textit{J. of Multivar. Analysis}, 42(2):171--201,
(1992).

\item {[32]} \textsc{V.B. Braginski, Y.I. Vorontzov, F.J. Halili}, \textit{%
Sov. Phys.--JETP}, 46(2):765, 4:171--201, (1977).

\item {[33]} \textsc{K.S. Thorne, R.W.P. Drever, C.M. Caves, M. Zimmermann,
V.D. Sandberg}, \textit{Phys. Rev. Lett.}, 40:667, (1978).

\item {[34]} \textsc{A.S. Holevo}, Quantum estimation. In Advances in
statistical signal processing, 1:157--202, (1987).

\item {[35]} \textsc{V.P. Belavkin}, Reconstruction theorem for quantum
stochastic processes, \textit{Theoret. Math. Phys.}, 3:409--431, (1985).

\item {[36]} \textsc{K. Kraus}, States, Effects and operations,
Springer--Verlag, Berlin 1983.

\item {[37]} \textsc{E.B. Ozawa}, \textit{J. Math. Phys.}, 25:79--87, (1984).

\item {[38]} \textsc{A. Barchielli, G. Lupieri}, \textit{J. Math. Phys.},
26:2222--2230, (1985).

\item {[39]} \textsc{L. Accardi, R. Alicki, A. Frigerio, Y.G. Lu}, An
invitation to weak coupling and low density limits, Quantum probability and
re. topics VI, ed. L. Accardi, 3--62, World Scientific, Singapore 1991.

\item {[40]} \textsc{P. Busch, P.J. Lahti, P. Mittelstaedt}, The quantum
theory of measurement, \textit{Lecture Notes in Physics}, Springer--Verlag,
Berlin 1991.
\end{description}

\end{document}